\documentclass[prb,twocolumn,10pt,aps,showpacs,floatfix,amsmath,amsfonts,letterpaper,longbibliography,nofootinbib]{revtex4-2}

\usepackage{stix} 
\usepackage{graphicx}
\usepackage{amsmath}
\usepackage{url}
\usepackage{array}
\usepackage{multirow}
\usepackage{bm}
\usepackage{mathtools}

\newcolumntype{C}{>{\centering\arraybackslash}p{2.5cm}}
\newcolumntype{P}[1]{>{\centering\arraybackslash}p{#1}}

\makeatletter
\def\beq{\@ifstar{\@ifnextchar[{\@beqslabel}{\@beqsnolabel}}
{\@ifnextchar[{\@beqlabel}{\@beqnolabel}}}
\def\@beqlabel[#1]{\begin{equation}\label{#1}}
\def\@beqnolabel{\begin{equation}}
\def\@beqslabel[#1]{\begin{equation*}\label{#1}}
\def\@beqsnolabel{\begin{equation*}}
\def\eeq{\@ifstar{\end{equation*}}{\end{equation}}}
\makeatother

\newcommand{\refcite}[1]{Ref.~\cite{#1}}
\newcommand{\refcites}[1]{Refs.~\cite{#1}}

\newcommand{\refeqq}[1]{Eq.~(\ref{#1})}
\newcommand{\refeqqs}[2]{Eqs.~(\ref{#1})\nobreakdash--(\ref{#2})}
\newcommand{\refeqqand}[2]{Eqs.~(\ref{#1}) and (\ref{#2})}

\newcommand{\reffig}[1]{Fig.~\ref{#1}}

\newcommand{\refsec}[1]{Sec.~\ref{#1}}
\newcommand{\refapp}[1]{Appendix~\ref{#1}}

\newcommand{\reftab}[1]{Table~\ref{#1}}

\newcommand{\reffootnote}[1]{Footnote~\ref{#1}}

\newcommand{\etal}{\textit{et al.}}

\newcommand{\ee}{\mathrm{e}}
\newcommand{\ii}{\mathrm{i}}

\newcommand{\punc}[1]{\,{\text{#1}}}
\newcommand{\sub}[1]{_{\text{#1}}}

\newcommand{\rv}{\bm{r}}
\newcommand{\Rv}{\bm{R}}
\newcommand{\deltav}{\bm{\delta}}
\newcommand{\goC}{\mathfrak{C}}

\newcommand{\ham}{\mathcal{H}}

\newcommand{\nF}{n\sub{F}}
\newcommand{\sigmax}{\bm{\sigma}^x}
\newcommand{\cd}{\bar{d}} 
\newcommand{\cdb}{\bar{\bm{d}}}
\newcommand{\thetav}{\bm{\theta}}

\DeclareMathOperator{\Tr}{Tr}
\DeclareMathOperator{\sgn}{sgn}

\DeclareMathOperator{\diag}{diag}
\DeclarePairedDelimiter\ceil{\lceil}{\rceil}
\DeclarePairedDelimiter\floor{\lfloor}{\rfloor}

\usepackage[usenames,dvipsnames]{color}

\begin{document}

\title{Topological sectors, dimer correlations and monomers from the transfer-matrix solution of the dimer model}

\author{Neil Wilkins}
\affiliation{School of Physics and Astronomy, The University of Nottingham, Nottingham, NG7 2RD, United Kingdom}

\author{Stephen Powell}
\affiliation{School of Physics and Astronomy, The University of Nottingham, Nottingham, NG7 2RD, United Kingdom}

\begin{abstract}
We solve the classical square-lattice dimer model with periodic boundaries and in the presence of a field $\bm{t}$ that couples to the (vector) flux, by diagonalizing a modified version of Lieb's transfer matrix. After deriving the torus partition function in the thermodynamic limit, we show how the configuration space divides into `topological sectors' corresponding to distinct values of the flux. Additionally, we demonstrate in general that expectation values are $\bm{t}$-independent at leading order, and obtain explicit expressions for dimer occupation numbers, dimer--dimer correlation functions and the monomer distribution function. The last of these is expressed as a Toeplitz determinant, whose asymptotic behavior for large monomer separation is tractable using the Fisher--Hartwig conjecture. Our results reproduce those previously obtained using Pfaffian techniques.
\end{abstract}

\maketitle

\section{Introduction}

The dimer model is a paradigmatic example of a strongly-correlated system, in which dimers cover the edges of a lattice subject to a close-packing constraint, i.e., each vertex touches exactly one dimer. It was first solved independently by Kasteleyn \cite{Kasteleyn1961,Kasteleyn1963} and Temperley and Fisher \cite{Temperley1961,Fisher1961} in 1961 using a combinatoric method, in which the partition function is expressed as the Pfaffian of a signed adjacency matrix known as the Kasteleyn matrix.

Because the dimer model is exactly solvable, it offers a useful setting for the study of novel phenomena in geometrically frustrated systems \cite{Balents2010} or, more specifically, `Coulomb-phase' physics \cite{Henley2010}. In particular, its extensive entropy reflects macroscopic ground-state degeneracy, while the configuration space splits into topological sectors labeled by horizontal and vertical `flux' components, reflecting topological order \cite{Castelnovo2012}. The Pfaffian method can be used to calculate partial partition functions for these sectors, as demonstrated by Boutillier and de Tili\`{e}re \cite{Boutillier2009}.

Moreover, a dimer can be replaced by a pair of monomers, which can be separated by subsequent dimer updates and thus play the role of fractionalized excitations. Fisher and Stephenson's Pfaffian calculation of the monomer distribution function in 1963 \cite{Fisher1963} implies that, due to the entropy of the background dimer configuration, the monomers interact through an effective Coulomb potential, which is logarithmic in two dimensions. They have also shown that dimer--dimer correlations are long-range with algebraic, rather than exponential, dependence on separation. This is despite the absence of long-range order, and instead a consequence of the close-packing constraint.

Perhaps a more elegant solution of the dimer model is Lieb's transfer-matrix method \cite{Lieb1967}, analogous to the well-known solution of the Ising model by Schultz \etal\ \cite{Schultz1964}, which maps the problem to free fermions. In this approach, the partition function is expressed in terms of a transfer matrix, which, given a configuration on a row of vertical bonds, generates all dimer configurations compatible with the close-packing constraint on the subsequent row of horizontal and vertical bonds. This can be expressed in terms of spin-$\frac{1}{2}$ operators and mapped to fermions through a Jordan--Wigner transformation.

This method has been used in the literature to derive the partition function \cite{Lieb1967} and to determine its vertical-flux decomposition \cite{Rasmussen2012, MorinDuchesne2016}. In this work, we show how Lieb's transfer matrix can be modified in order to calculate the full flux-sector decomposition. We also provide a general framework for the calculation of expectation values and explicitly calculate dimer occupation numbers, dimer--dimer correlation functions and the monomer distribution function. For the last of these, we show how the asymptotic dependence for large monomer separation, which was deduced by numerical means in \refcite{Fisher1963}, can be evaluated exactly by applying the Fisher--Hartwig conjecture \cite{Fisher1969}.

\subsection*{Outline}

In \refsec{model} we define the model before showing how it can be formulated in terms of a transfer matrix in \refsec{transfermatrix}. We then diagonalize the two-row transfer matrix in \refsec{diagonalizationofthetworowtransfermatrix}, whose spectrum is used to calculate the partition function, including its flux-sector decomposition, in \refsec{partitionfunction}, and various expectation values in \refsec{expectationvalues}. We conclude in \refsec{conclusions}.

\section{Model}
\label{model}

We consider the close-packed dimer model on an $L_{x} \times L_{y}$ square lattice with periodic boundary conditions (PBCs), assuming both $L_{x}$, $L_{y}$ even. In the following, we define the flux along with the weights that appear in the partition function.

Denoting by $d_{\rv,\mu}$ the dimer occupation number (equal to zero or one) on the bond joining sites $\rv$ and $\rv + \deltav_\mu$, with \(\deltav_\mu\) a unit vector in direction $\mu \in \{x,y\}$, the flux is given by
\begin{equation}
\label{eq:fluxdef}
\Phi_{\mu} = \frac{1}{L_{\mu}}\sum_{\rv}\epsilon_{\bm{r}}d_{\rv,\mu} \punc,
\end{equation}
where $\epsilon_{\rv} = (-1)^{r_x + r_y} = \pm 1$ depending on the sublattice. Due to the close-packing constraint, this is equivalent to the sum of $\epsilon_{\bm{r}}d_{\rv,\mu}$ on links crossing a surface normal to $\deltav_{\mu}$ (to see this, one usually defines an effective `magnetic field' \cite{Huse2003,Henley2010}). The latter definition highlights that $\Phi_{\mu}$ is integer valued, and can only be changed by shifting dimers around a loop encircling the whole system \cite{Chalker2017}. The flux thus plays the role of a topological invariant.

\begin{figure}
\begin{center}
\includegraphics[width=0.85\columnwidth]{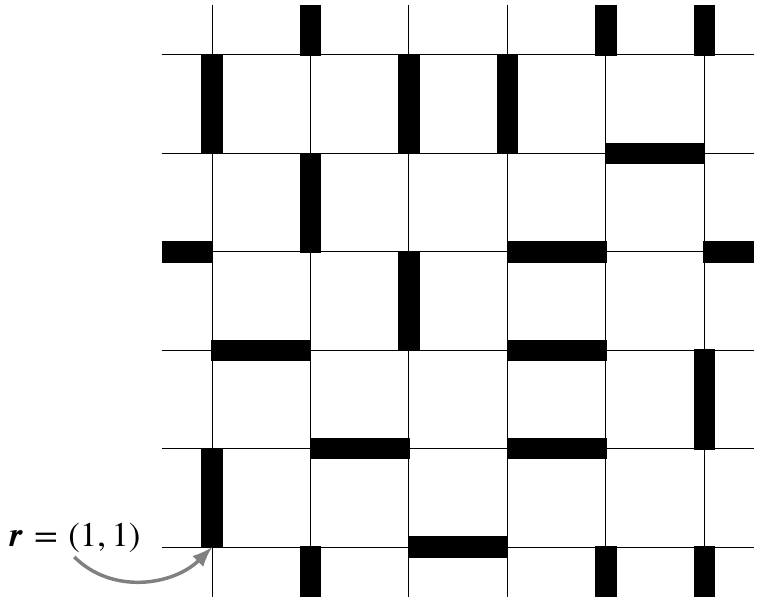}
\caption{An example configuration of the close-packed dimer model on a $6\times 6$ lattice with periodic boundaries. The number of horizontal dimers is $N_{x}=8$ and the flux is $\bm{\Phi}=(1,1)$ [see \refeqq{eq:fluxdef} and text thereafter]. Hence, this configuration has weight $\alpha^{8}\ee^{\ii \bm{t}\cdot (1,1)}$.}
\label{fig:model}
\end{center}
\end{figure}

To each configuration, we assign weight $\alpha^{N_{x}}\ee^{\ii\bm{t}\cdot\bm{\Phi}}$. In the first factor $\alpha > 0$ and $N_{x}$ are the `activity' and number of horizontal dimers, respectively. (The total number of dimers \(N_x+N_y = \frac{1}{2}L_xL_y\) is fixed, so the activity of vertical dimers is set to unity without loss of generality.) Hence, for $\alpha \neq 1$, the model is anisotropic, with horizontal (vertical) dimers favored for $\alpha >1$ ($\alpha < 1$). In the second factor $\bm{t}$ is a field, with components $t_{\mu} \in (-\pi, \pi]$, that couples to the flux $\bm{\Phi}$. An example configuration is shown in \reffig{fig:model}.

The partition function is
\begin{equation}
\label{eq:partitionfunction}
Z(\bm{t}) = \sum_{c\in\goC_0} \alpha^{N_{x}}\ee^{\ii\bm{t}\cdot\bm{\Phi}} \punc ,
\end{equation}
where $\goC_0$ denotes the set of all close-packed dimer configurations, and can be thought of as a moment-generating function for $\Phi_{\mu}$. Similarly, expectation values of a function $O$ of the dimer occupation numbers $d_{\rv,\mu}$ are given by
\begin{equation}
\label{eq:expectationvalue}
\langle O \rangle = \frac{1}{Z(\bm{t})}\sum_{c\in\goC_0} O\alpha^{N_{x}}\ee^{\ii\bm{t}\cdot\bm{\Phi}} \punc .
\end{equation}

\section{Transfer matrix}
\label{transfermatrix}

We construct the partition function, \refeqq{eq:partitionfunction}, by modifying Lieb's transfer matrix \cite{Lieb1967} to include the $\Phi_{x}$ weighting (the $\Phi_{y}$ weighting can be included without modifying the transfer matrix).

\begin{figure*}
\begin{center}
\includegraphics[width=0.9\textwidth]{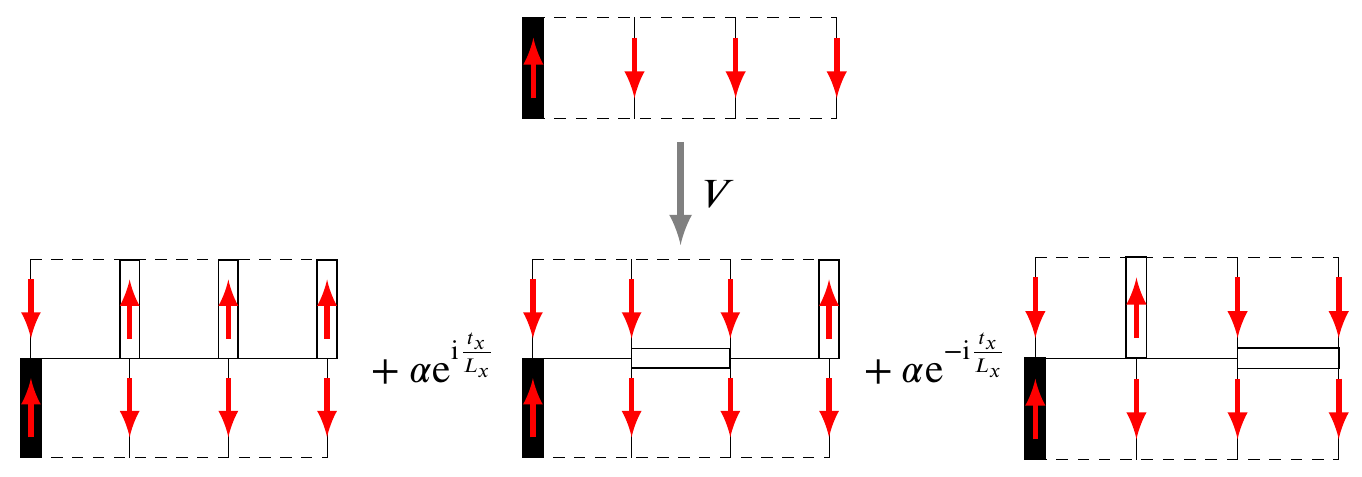}
\caption{Action of the transfer matrix $V$ of \refeqq{eq:tm} on a row of vertical bonds (top), in which occupied and empty vertical bonds are represented by spin up $\lvert \uparrow \rangle$ and down $\lvert \downarrow \rangle$ states (red), respectively. The result is all dimer configurations on the subsequent row of vertical bonds that are consistent with the close-packing constraint (bottom). The left configuration with all dimers vertical is generated by $V_{1}$, which flips all spins. The middle and right configurations, obtained from the left configuration by replacing pairs of neighboring vertical dimers with horizontal dimers, are generated by $V_{3}$, which flips neighboring up spins. In order to obtain the correct weights in the partition function, \refeqq{eq:partitionfunction}, $V$ and $V^{*}=V^{\dagger}$ act on alternate rows and assign weight $\mu_{j}=\alpha\exp{[\ii t_{x}(-1)^{j}/L_{x}]}$ and $\mu_{j}^{*}$ to a horizontal dimer between sites $j$ and $j+1$, respectively.}
\label{fig:transfermatrixspins}
\end{center}
\end{figure*}

We first define a vector space whose basis vectors \(\lvert \cdb_y\rangle\) correspond to all possible configurations \(\cdb_y\) of the dimer occupation numbers on a single row of vertical bonds. As illustrated in \reffig{fig:transfermatrixspins}, the transfer matrix \(V\) is defined so that
\beq[EqVdket]
V\lvert \cdb_y\rangle = \sum_{\cdb_y'}\lvert\cdb_y'\rangle\sum_{\cdb_x \in \goC(\cdb_y,\cdb_y')}  w(\cdb_x)\punc,
\eeq
where \(\cdb_y'\) is the configuration on the subsequent row of vertical bonds and \(\goC(\cdb_y,\cdb_y')\) is the (possibly empty) set of configurations $\cdb_x$ of the intermediate row of horizontal bonds that are compatible with \(\cdb_y\) and \(\cdb_y'\). The weight function $w$ is chosen to give the correct weights for $N_{x}$ and $\Phi_{x}$ in the partition function of \refeqq{eq:partitionfunction}. On even rows, where \(\epsilon_{\rv} = (-1)^{r_x}\) in \refeqq{eq:fluxdef}, it is given by
\beq[eq:wd]
w(\cdb_x) = \prod_{j=1}^{L_x} \mu_j^{ \cd_{j,x}}
\punc,
\eeq
where
\begin{equation}
\label{eq:muj}
\mu_{j} = \alpha \exp\left[\ii (-1)^{j}\frac{t_{x}}{L_{x}}\right] \punc ,
\end{equation}
while on odd rows \(w\) is defined in the same way, but with \(\mu_j\) replaced by \(\mu_j^*\). (Here, \(\cd_{j,x}\) denotes the occupation number of the bond between sites \(r_x=j\) and \(j+1\) in the configuration \(\cdb_x\) of the horizontal bonds.)

It is convenient to split the action of \(V\) into two steps:
\begin{enumerate}
\item Generate the (single) configuration \(\cdb_y'= (1,1,\dotsc,1)-\cdb_y\) with all horizontal bonds on the intermediate row empty (left configuration in \reffig{fig:transfermatrixspins}).
\item Starting with the result of step 1, one may produce all other configurations by replacing pairs of neighboring vertical dimers with a horizontal dimer (middle and right configurations in \reffig{fig:transfermatrixspins}). The effect on \(\cdb_y'\) is that an adjacent pair of dimers is removed.
\end{enumerate}
In order to reproduce the weight function \(w\), a horizontal dimer on the bond between sites \(j\) and \(j+1\) in step 2 comes with a factor $\mu_j$ ($\mu_j^*$) on even (odd) rows.

An explicit operator expression for the transfer matrix is obtained by representing occupied and empty vertical bonds by spin up $\lvert \uparrow \rangle$ and down $\lvert \downarrow \rangle$ states, respectively [i.e., eigenstates of $\sigma_{j}^{z}$, where $\bm{\sigma}_{j}=(\sigma_{j}^{x},\sigma_{j}^{y},\sigma_{j}^{z})$ are the Pauli matrices]. The above steps are easy to formulate in the spin language. As shown in \reffig{fig:transfermatrixspins}, step 1 is equivalent to flipping all spins, which is achieved by the operator
\begin{equation}
V_{1} = \prod_{j=1}^{L_{x}}\sigma_{j}^{x} \punc ,
\end{equation}
since $\sigma_{j}^{\pm} = \frac{1}{2}(\sigma_{j}^{x} \pm i \sigma_{j}^{y})$ satisfy $\sigma^{+}\lvert \downarrow \rangle = \lvert \uparrow \rangle$ and $\sigma^{-}\lvert \uparrow \rangle = \lvert \downarrow \rangle$. 

In step 2, pairs of neighboring up spins are flipped, so the operator
\begin{equation}
\label{eq:nxspin}
d_{j,x} = \mu_{j}\sigma_{j}^{-}\sigma_{j+1}^{-}
\end{equation}
effectively generates a horizontal dimer between sites $j$ and $j+1$, with the correct weight on even rows. Because $(\sigma_{j}^{-})^{2} = 0$, the operator $(m!)^{-1}\left(\sum_{j=1}^{L_{x}}d_{j,x}\right)^{m}$ generates $m$ horizontal dimers (PBCs require $\sigma^{-}_{L_{x} + 1} = \sigma^{-}_{1}$), and hence
\begin{equation}
\label{eq:V3}
V_{3} = \exp\left(\sum_{j=1}^{L_{x}}d_{j,x}\right)
\end{equation}
generates an arbitrary number of horizontal dimers. To obtain the correct weights on odd rows, one should instead use the operator $V_{3}^{*}$.

It is therefore necessary to define two transfer matrices,
\begin{equation}
\label{eq:tm}
V = V_{3}V_{1}
\end{equation}
on even rows and $V^{*}=V^\dagger$ on odd rows.\footnote{Lieb's transfer matrix $V = V_{3}V_{2}V_{1}$ includes a third operator $V_{2}$, which generates an arbitrary number of monomers on a row \cite{Lieb1967}.} (Note that \(V^T = V\) because $\sigma^{x}\sigma^{+} = \sigma^{-}\sigma^{x}$.) We also define the two-row transfer matrix
\begin{equation}
\label{eq:tmstart}
W = VV^\dagger = V_{3}V_{3}^{\dagger} \punc ,
\end{equation}
which is manifestly Hermitian.

The $\Phi_{y}$ weighting is included in the transfer-matrix formalism as follows: The operator for the dimer occupation number on a vertical bond is simply
\begin{equation}
\label{eq:nyspin}
d_{j,y} = \frac{1}{2}(1 + \sigma_{j}^{z}) \punc ,
\end{equation}
since spin up (down) corresponds to an occupied (empty) bond. In terms of this, the vertical flux component on even rows is [see \refeqq{eq:fluxdef} and text thereafter]
\begin{equation}
\label{eq:yflux}
\Phi_{y} = \sum_{j=1}^{L_{x}}(-1)^{j}d_{j,y} \punc ,
\end{equation}
which satisfies the (anti)commutation relations $\{\Phi_{y},V\} = 0$ and $[\Phi_{y}, W] = 0$.\footnote{$\Phi_{y}$ appears in \refcites{Rasmussen2012,MorinDuchesne2015,MorinDuchesne2016} as the operator $\mathcal{V}$, whose eigenvalues are referred to as the `variation index'.} The latter implies that it is possible to construct mutual eigenstates of the two-row transfer matrix $W$ and $\Phi_{y}$. The partition function, \refeqq{eq:partitionfunction}, is then given by
\begin{equation}
\label{eq:Ztrace}
Z(\bm{t}) = \Tr\left[\ee^{\ii t_{y}\Phi_{y}}W^{\frac{L_{y}}{2}}\right]
\end{equation}
(the trace arises due to PBCs in the vertical direction).

Similarly, the operator analog of \refeqq{eq:expectationvalue}, in the case of the correlation function between observables $O$ and $O'$ in rows \(1 \le l \le l' \le L_y\), is given by
\begin{equation}
\label{eq:correlatortrace}
\langle O'(l')O(l) \rangle = \frac{1}{Z(\bm{t})}\Tr\left[\ee^{\ii t_{y}\Phi_{y}}W^{\frac{L_{y}}{2}}O'(l')O(l)\right] \punc ,
\end{equation}
where \(O(l) = U(l)^{-1}O U(l)\) and
\beq[eq:Ul]
U(l) = \underbrace{\dotsm V^\dagger V V^\dagger}_l = 
\begin{cases}
V^\dagger W^{(l-1)/2} & \text{for $l$ odd} \\
W^{l/2} & \text{\phantom{for} $l$ even.}
\end{cases}
\eeq
Note that \([O(l)]^\dagger = O^\dagger(-l)\), where \(U(-l)=[U(l)^\dagger]^{-1}\) is defined by the second equality of \refeqq{eq:Ul}.

To compute expectation values of dimer observables, it is necessary to find operators that correspond to these quantities. While a suitable operator for the dimer occupation number on vertical bonds has already been defined in \refeqq{eq:nyspin}, no such operator exactly represents the dimer occupation number on horizontal bonds, since the vector space on which the transfer matrix acts contains only dimer configurations on vertical bonds.

One can nonetheless calculate expectation values involving horizontal dimers using an appropriately constructed operator. From \refeqqand{EqVdket}{eq:wd}, one finds
\beq
\mu_j \frac{\partial}{\partial \mu_j}V\lvert \cdb_y\rangle = \sum_{\cdb_y'}\lvert\cdb'_y\rangle\sum_{\cdb_x \in \goC(\cdb_y,\cdb_y')} \cd_{j,x}w(\cdb_x)\punc,
\eeq
whereas \refeqqs{eq:nxspin}{eq:tm} give the operator identity
\beq
\mu_j \frac{\partial}{\partial \mu_j} V = d_{j,x} V\punc,
\eeq
since \([d_{j,x},d_{j',x}]=0\). Comparing the right-hand sides, we therefore interpret \(d_{j,x}\) as the operator corresponding to the horizontal dimer occupation number \(\cd_{j,x}\) on an even row, but only when appearing in the combination\footnote{\label{note:djx}This means that, for example, \(d_{j,x}^2\) does not give the square of the horizontal dimer number; in fact \(d_{j,x}^2 = 0\), whereas \(\cd_{j,x}^2 = \cd_{j,x}\).} \(d_{j,x} V\). Similarly, $d_{j,x}^{*}$ acts as the horizontal dimer occupation number on an odd row in the combination \(d_{j,x}^* V^\dagger\). Setting \(O\) equal to \(d_{j,x}\) (\(d_{j,x}^*\)) on even (odd) rows in \refeqq{eq:correlatortrace} gives the correct combination \(d_{j,x}V\) (\(d_{j,x}^*V^\dagger\)) in $O(l)$, allowing one to calculate expectation values involving the horizontal dimer number.

\section{Diagonalization of the two-row transfer matrix}
\label{diagonalizationofthetworowtransfermatrix}

To calculate \refeqq{eq:Ztrace} it is sufficient to diagonalize the two-row transfer matrix $W$. We do so in this section through a series of transformations.

We map between spins and spinless fermions using the Jordan--Wigner transformation \cite{Jordan1928,Lieb1961,Sachdev2011}
\begin{align}
\label{eq:JW1}
&C_{j} = \left( \prod_{i=1}^{j-1}- \sigma_{i}^{z} \right)\sigma_{j}^{-} \\
\label{eq:JW2}
&C_{j}^{\dagger} = \left( \prod_{i=1}^{j-1}- \sigma_{i}^{z} \right)\sigma_{j}^{+} \\
\label{eq:JW3}
&C_{j}^{\dagger}C_{j} = \frac{1}{2}(1 + \sigma_{j}^{z}) \punc ,
\end{align}
which identifies spin up and down with filled and empty fermion orbitals, respectively, while preserving the usual (anti)commutation relations
\begin{gather}
[\sigma_{i}^{\mu},\sigma_{j}^{\nu}] = 2i\delta_{ij}\epsilon_{\mu\nu\rho}\sigma_{\rho} \\
\label{eq:fermcom}
\phantom{\punc .}\{C_{i}, C_{j} \} = \{C_{i}^{\dagger}, C_{j}^{\dagger} \} = 0 \qquad \{C_{i}, C_{j}^{\dagger}\} = \delta_{ij} \punc .
\end{gather}

In terms of fermions, \refeqqand{eq:nxspin}{eq:nyspin} become
\begin{align}
\label{eq:dxc}
d_{j,x} &= -\mu_{j}C_{j}C_{j+1}\\
\label{eq:dyc}
d_{j,y} &= C_{j}^{\dagger}C_{j}
\punc ,
\end{align}
while the condition $\sigma_{L_{x}+1}^{-}=\sigma_{1}^{-}$ is equivalent to 
\begin{equation}
\label{eq:cbcs1}
C_{L_{x}+1}=-C_{1}(-1)^{\Phi_{y}} = (-1)^{\Phi_{y}}C_{1}
\end{equation}
with
\begin{equation}
\label{eq:yflux1}
\Phi_{y} = \sum_{j} (-1)^{j}C_{j}^{\dagger}C_{j}\punc.
\end{equation}

We now define projectors
\beq[eq:definePi]
\Pi_p = \frac{1}{2}[1 + (-1)^p(-1)^{\Phi_y}]
\eeq
into the subspaces with even (\(p = 0\)) or odd (\(p = 1\)) \(\Phi_y\), which satisfy $\sum_{p}\Pi_{p} = 1$ and $(-1)^{\Phi_{y}}\Pi_{p} = (-1)^{p}\Pi_{p}$. Then, since $(-1)^{\Phi_{y}}$ commutes with any quadratic form in fermions, we have
\begin{align}
W &= W\sum_{p}\Pi_{p} \\
\label{eq:wprojectors}
&= \sum_{p}W_{p}\Pi_{p} \punc ,
\end{align}
where
\begin{equation}
\label{eq:tmjw}
W_{p} = \exp{\left(-\sum_{j=1}^{L_{x}} \mu_{j}C_{j}C_{j+1}\right)} \times \text{h.c.} \punc ,
\end{equation}
and the fermion operator $C_{L_{x}+1}$ depends implicitly on $p$ through the boundary condition
\begin{equation}
\label{eq:cbcs}
C_{L_{x} + 1} = -(-1)^{p}C_{1}.
\end{equation}

More generally, for any operator $O$ containing $C_{L_{x}+1}$ of \refeqq{eq:cbcs1}, we define an operator $O_{p}$ that instead only contains $C_{L_{x}+1}$ of \refeqq{eq:cbcs} (and thus depends on $p$), such that the action of both operators on a state with $\Phi_{y}$ parity $p$ yields the same result, i.e., $O = \sum_{p}O_{p}\Pi_{p}$. (For operators that do not contain $C_{L_{x}+1}$, such as $\Phi_{y}$, one has $O_{p}=O$.)

For later reference (see \refsec{expectationvalues}) we note that, after the Jordan--Wigner transformation, the single-row transfer matrix is given by \(V = \sum_p V_p \Pi_p\), with
\beq[eq:tmjv]
V_p = \exp{\left(-\sum_{j=1}^{L_{x}} \mu_{j}C_{j}C_{j+1}\right)} \prod_{j=1}^{L_x}\left[C_j + (-1)^j C_j^\dagger\right] \punc ,
\eeq
where the operators in the product should be ordered from right to left.

We now make a Fourier expansion
\begin{equation}
\label{eq:ceta}
C_{j} = \frac{\ee^{-\ii \pi/4}}{\sqrt{L_{x}}}\sum_{k \in \mathbb{K}_{p}}\ee^{\ii kj}\eta_{k} \punc ,
\end{equation}
with
\begin{equation}
\label{eq:qevenflux}
\mathbb{K}_{0} = \{\pm\pi/L_{x}, \pm 3\pi/L_{x}, \dotsc , \pm(L_{x}-1)\pi/L_{x}\}
\end{equation}
and
\begin{equation}
\label{eq:qoddflux}
\mathbb{K}_{1} = \{0, \pm 2\pi/L_{x}, \pm 4\pi/L_{x}, \dotsc , \pm(L_{x}-2)\pi/L_{x}, \pi\} \punc ,
\end{equation}
which ensure the correct boundary condition on $C_{L_{x}+1}$ in \refeqq{eq:cbcs} \cite{Lieb1967}.\footnote{\label{FootnoteTwistedBC}As an alternative to the approach in \refsec{transfermatrix}, one could instead implement the $\Phi_{x}$ weighting using $\mu_{j} = \alpha$ and twisted boundary conditions $\sigma^-_{L_{x}+1}=\ee^{\ii t_{x}}\sigma^-_{1}$ in place of \refeqq{eq:muj} and $\sigma^-_{L_{x}+1}=\sigma^-_{1}$, respectively [see \refeqq{eq:fluxdef} and text thereafter]. However, a Fourier expansion of the new set of fermions $\tilde{C}_j$ is no longer useful because of the absence of translation symmetry \cite{Cabrera1987,Abraham1988}. Instead, one would have to perform the gauge transformation \(\tilde{C}_j = \ee^{-\ii j(-1)^j t_x/L_x} C_j\) back to $C_{j}$ fermions, before proceeding as in the main text.} The $\eta_{k}$ fermions obey standard anticommutation relations, as follows from \refeqq{eq:fermcom}.

Using the result
\begin{multline}
\label{eq:result}
\frac{1}{L_{x}}\sum_{j=1}^{L_{x}}\mu_{j}\ee^{\ii (k+k')j} = {} \\ \alpha\left[ \delta_{k+k',0}\cos \left(\frac{t_{x}}{L_x}\right) + \ii \delta_{k+k',\pi}\sin \left(\frac{t_{x}}{L_x}\right)\right] \punc,
\end{multline}
valid for both \(k\) and \(k'\) in either \(\mathbb{K}_{0}\) or \(\mathbb{K}_{1}\), the operator appearing in the exponential of \refeqq{eq:tmjw} can be written as
\begin{multline}
-\sum_{j=1}^{L_{x}} \mu_{j}C_{j}C_{j+1} = {} \\ \ii \alpha \sum_{k\in \mathbb{K}_p}\ee^{-\ii k}\eta_{k}\left[\cos\left(\frac{t_{x}}{L_{x}}\right) \eta_{-k} - \ii \sin\left(\frac{t_{x}}{L_{x}}\right)\eta_{\pi - k}\right] \punc .
\end{multline}
Restricting the sum to $0\leq k \leq \frac{\pi}{2}$, this becomes
\beq[eq:tmftlemma]
-\sum_{j=1}^{L_{x}} \mu_{j}C_{j}C_{j+1} = \sum_{\substack{k\in\mathbb{K}_{p} \\ 0 \leq k \leq \frac{\pi}{2}}} Q_k\bm(\mathbf{A}(k)\bm)\punc ,
\eeq
where the quadratic form
\begin{equation}
\label{eq:defineQ}
Q_{k}(\mathbf{X}) =
\begin{cases}
\frac{1}{2}\bm{\eta}_{k}^{\dagger} \mathbf{X} \bm{\eta}_{k} & \text{for \(k \in \left\{0, \frac{\pi}{2}\right\}\)} \\
\bm{\eta}_{k}^{\dagger} \mathbf{X} \bm{\eta}_{k} & \text{otherwise.}
\end{cases}
\end{equation}
Here,
\begin{equation}
\bm{\eta}_{k} = 
\begin{pmatrix}
\eta_{k} \\
\eta_{k-\pi} \\
\eta_{-k}^{\dagger} \\
\eta_{\pi - k}^{\dagger}
\end{pmatrix}
\end{equation}
[its Hermitian conjugate means the row vector $\bm{\eta}_{k}^{\dagger} =
(\eta_{k}^{\dagger} \,\, \eta_{k-\pi}^{\dagger} \,\, \eta_{-k} \,\, \eta_{\pi - k})$], while the $4 \times 4$ matrix
\begin{equation}
\mathbf{A}(k) = 
\label{eq:A}
\begin{pmatrix}
0 & 0 \\
\mathbf{A}_{21} & 0
\end{pmatrix} \punc ,
\end{equation}
with
\begin{equation}
\mathbf{A}_{21} = 2\alpha 
\begin{bmatrix}
-\sin k\cos \left(\frac{t_{x}}{L_x}\right) & \cos k\sin \left(\frac{t_{x}}{L_x}\right) \\
-\cos k\sin \left(\frac{t_{x}}{L_x}\right) &  \sin k\cos\left(\frac{t_{x}}{L_x}\right)
\end{bmatrix} \punc .
\end{equation}
The additional factor of $\frac{1}{2}$ for $k\in\{0,\frac{\pi}{2}\}$ prevents double counting of these terms in \refeqq{eq:tmftlemma}, and ensures the commutation relation
\begin{equation}
\label{eq:qcomm}
\left[Q_{k}(\mathbf{X}), Q_{k'}(\mathbf{Y})\right] = \delta_{kk'}Q_{k}\left(\left[\mathbf{X},\mathbf{Y}\right]\right) \punc ,
\end{equation}
is valid for all \(0 \le k \le \frac{\pi}{2}\).\footnote{For $k\in\{0,\frac{\pi}{2}\}$, because of the nonzero anticommutator $\{\eta_{k,i},\eta_{k,j}\} = (\mathbf{W}_{k})_{i,j}$,
where
\begin{align}
\mathbf{W}_{0} = \sigmax\otimes I_2 &&& \mathbf{W}_{\pi/2} = \sigmax\otimes \sigmax\punc,
\end{align}
with \(\otimes\) denoting the Kronecker product,
\refeqq{eq:qcomm} is only true if $\mathbf{X}$ satisfies the condition $\mathbf{W}_{k}\mathbf{X}^{T}\mathbf{W}_{k} = -\mathbf{X}$ (or the same for $\mathbf{Y}$). However, it is always possible to symmetrize $\mathbf{X}$ to meet this condition: Using \(
(\bm{\eta}_{k}^\dagger){}^T = \mathbf{W}_{k}\bm{\eta}_{k,j}\), one can show
\begin{equation}
Q_{k}(\mathbf{X}) = Q_{k}(\mathbf{X}') + \frac{1}{2}\Tr(\mathbf{W}_{k}\mathbf{X}^{T}\mathbf{W}_{k}) \punc ,
\end{equation}
where $\mathbf{X}'=\frac{1}{2}(\mathbf{X} - \mathbf{W}_{k}\mathbf{X}^{T}\mathbf{W}_{k})$ is a matrix that satisfies the condition. The matrix $\mathbf{A}(k)$ in \refeqq{eq:A} has been constructed in this way.}

Since \(Q_k^\dagger(\mathbf{X}) = Q_k(\mathbf{X}^\dagger)\), and all quadratic forms in \refeqq{eq:tmftlemma} commute by \refeqq{eq:qcomm}, the two-row transfer matrix, \refeqq{eq:tmjw}, is given by
\begin{equation}
W_{p} =
\left[\array{@{}c@{}}
  \raisebox{-2ex}{$\displaystyle\prod$} \\
  \raisebox{-5.5pt}{$\substack{k\in\mathbb{K}_{p} \\ 0 \leq k \leq \frac{\pi}{2}}$}\endarray 
\ee^{Q_k\bm(\mathbf{A}(k)\bm)}\right] \left[\array{@{}c@{}}
  \raisebox{-2ex}{$\displaystyle\prod$} \\
  \raisebox{-5.5pt}{$\substack{k\in\mathbb{K}_{p} \\ 0 \leq k \leq \frac{\pi}{2}}$}\endarray 
\ee^{Q_k\bm(\mathbf{A}^{\dagger}(k)\bm)}\right] \punc ,
\end{equation}
which can be reordered as the following product of commuting terms:
\begin{equation}
\label{eq:tmft}
W_{p} = \prod_{\substack{k\in\mathbb{K}_{p} \\ 0 \leq k \leq \frac{\pi}{2}}} \ee^{Q_{k}\bm(\mathbf{A}(k)\bm)}\ee^{Q_{k}\bm(\mathbf{A}^{\dagger}(k)\bm)} \punc .
\end{equation}

To proceed, we map to the corresponding one-dimensional quantum Hamiltonian $\ham$ through
\begin{equation}
\label{eq:ham}
W = \ee^{-2\ham} \punc .
\end{equation}
Then, by \refeqq{eq:wprojectors}, we have
\beq[eq:hamprojected]
\ham = \sum_{p}\ham_{p}\Pi_{p} \punc ,
\eeq
where
\begin{equation}
\label{eq:wphp}
W_{p} = \ee^{-2\ham_{p}} \punc ,
\end{equation}
since the projectors satisfy $[\Pi_{p},W_{p'}]=0$ and $\Pi_{p}\Pi_{p'}=\Pi_{p}\delta_{pp'}$. After inserting \refeqq{eq:tmft}, this implies
\begin{equation}
\label{eq:hamp}
\ham_{p} = -\frac{1}{2}\sum_{\substack{k\in\mathbb{K}_{p} \\ 0 \leq k \leq \frac{\pi}{2}}}\log \left[ \ee^{Q_{k}\bm(\mathbf{A}(k)\bm)}\ee^{Q_{k}\bm(\mathbf{A}^{\dagger}(k)\bm)} \right] \punc .
\end{equation}

The Baker--Campbell--Hausdorff formula \cite{Rossmann2006} states that the logarithm in \refeqq{eq:hamp} can be expressed in terms of nested commutators of \(Q_k(\mathbf{A})\) and \(Q_k(\mathbf{A}^\dagger)\). Using \refeqq{eq:qcomm}, these can be expressed in terms of nested commutators of \(\mathbf{A}\) and \(\mathbf{A}^\dagger\), giving
\begin{equation}
\label{eq:ham1}
\ham_{p} = -\frac{1}{2}\sum_{\substack{k\in\mathbb{K}_{p} \\ 0 \leq k \leq \frac{\pi}{2}}} Q_{k} \left( \log ( \ee^{\mathbf{A}(k)}\ee^{\mathbf{A}^{\dagger}(k)} ) \right ) \punc .
\end{equation}
The problem is thus reduced to diagonalization of the $4 \times 4$ matrix $\ee^{\mathbf{A}}\ee^{\mathbf{A}^{\dagger}}$ for each $k$.

In order to solve the eigenvalue problem
\begin{equation}
\label{eq:eigprob}
\ee^{\mathbf{A}}\ee^{\mathbf{A}^{\dagger}}\bm{v} = \lambda\bm{v} \punc ,
\end{equation}
we expand $\ee^{\mathbf{A}}$ as a power series and use $\mathbf{A}^{2} = 0$ to obtain
\begin{equation}
\ee^{\mathbf{A}}\ee^{\mathbf{A}^{\dagger}} = \mathbf{I} + \mathbf{A} + \mathbf{A}^{\dagger} + \mathbf{A}\mathbf{A}^{\dagger} \punc .
\end{equation}
After substituting \refeqq{eq:A} and writing $\bm{v} = (\bm{v}_{1} \,\, \bm{v}_{2})^{T}$, \refeqq{eq:eigprob} reduces to a pair of simultaneous equations which, on rearrangement, read
\begin{align}
\bm{v}_{1} &= \frac{1}{\lambda - 1}\mathbf{A}_{21}^{\dagger}\bm{v}_{2} \\
\label{eq:eig22}
\mathbf{A}_{21}\mathbf{A}_{21}^{\dagger}\bm{v}_{2} &= \frac{(\lambda - 1)^{2}}{\lambda}\bm{v}_{2} \punc .
\end{align}
The latter is a $2 \times 2$ eigenvalue problem, which is easily solved. The result implies
\begin{multline}
\label{eq:diag}
\ee^{\mathbf{A}}\ee^{\mathbf{A}^{\dagger}} =  \mathbf{U}\diag[\lambda_{-}(k-t_{x}/L_{x}),\lambda_{+}(k-t_{x}/L_{x}), {} \\ \lambda_{-}(k + t_{x}/L_{x}),\lambda_{+}(k + t_{x}/L_{x})]\mathbf{U}^{\dagger} \punc ,
\end{multline}
where
\begin{equation}
\lambda_{\pm}(k) = \left[\alpha\sin k \pm (1 + \alpha^{2}\sin ^{2} k)^{\frac{1}{2}}\right]^{2} \punc ,
\end{equation}
and $\mathbf{U}$ is a unitary matrix whose columns are the eigenvectors of $\ee^{\mathbf{A}}\ee^{\mathbf{A}^{\dagger}}$.

By inserting \refeqq{eq:diag} into \refeqq{eq:ham1}, we obtain the free-fermion Hamiltonian
\begin{equation}
\label{eq:ham2}
\ham_{p} = \sum_{k\in\mathbb{K}_{p}}\epsilon(k-t_{x}/L_{x})\zeta_{k}^{\dagger}\zeta_{k} \punc ,
\end{equation}
with dispersion
\begin{equation}
\label{eq:dispersion}
\epsilon(k) = \frac{1}{2}\log{\lambda_{+}}(k) = \sinh^{-1}(\alpha\sin k) \punc,
\end{equation}
where the $\zeta_{k}$ and $\eta_{k}$ fermions are related by the Bogoliubov transformation
\begin{equation}
\label{eq:bog0}
\bm{\zeta}_{k} =
\begin{pmatrix}
\zeta_{k} \\
\zeta_{k-\pi} \\
\zeta_{-k}^{\dagger} \\
\zeta_{\pi - k}^{\dagger}
\end{pmatrix} = \mathbf{U}^{\dagger}\bm{\eta}_{k} \punc ,
\end{equation}
for $0 \leq k \leq \pi/2$. Both sets of fermions obey standard anticommutation relations.

The transformation of \refeqq{eq:bog0} may be expressed as a single transformation valid for all $k$:
\begin{multline}
\label{eq:bog}
\eta_{k} = \frac{1}{\sqrt{2}}\Big(\cos\theta_{k-t_{x}/L_{x}}\zeta_{k} + \cos\theta_{k+t_{x}/L_{x}}\zeta_{-k}^{\dagger} - {} \\ \sin\theta_{k+t_{x}/L_{x}}\zeta_{\pi - k}^{\dagger} + \sin\theta_{k-t_{x}/L_{x}}\zeta_{k - \pi} \Big) \punc ,
\end{multline}
with
\begin{equation}
\label{eq:angle}
\tan(2\theta_{k}) = \frac{1}{\alpha \sin k} , \qquad \theta_{k} \in \left[0 ,\frac{\pi}{2}\right] \punc .
\end{equation}
Combining \refeqqand{eq:ceta}{eq:bog}, the transformation relating the $C_{j}$ and $\zeta_{k}$ fermions is
\begin{equation}
\label{eq:czeta}
C_{j} =
\sqrt{\frac{2}{L_{x}}}\ee^{-\ii \pi/4}\sum_{k\in\mathbb{K}_{p}}\ee^{\ii kj} \times \begin{cases}\cos\theta_{k+t_x/L_x}\zeta_{-k}^{\dagger} & \text{for \(j\) odd} \\
\cos\theta_{k-t_x/L_x}\zeta_{k} & \text{\phantom{for} \(j\) even,}
\end{cases}
\end{equation}
with inverse
\begin{multline}
\zeta_k = \sqrt{\frac{2}{L_x}}\ee^{\ii \pi / 4} \cos\theta_{k-t_{x}/L_x}\sum_{\text{even \(j\)}} \ee^{-\ii k j} C_j + {} \\ \sqrt{\frac{2}{L_x}}\ee^{-\ii \pi / 4} \sin\theta_{k-t_{x}/L_x}\sum_{\text{odd \(j\)}} \ee^{-\ii k j} C_j^\dagger\punc.
\end{multline}
This makes it clear that the annihilation operator \(\zeta_k\) removes a fermion (or equivalently, removes a vertical dimer) on even sites or adds one on odd sites. According to \refeqq{eq:yflux1}, it therefore reduces \(\Phi_y\) by one.

We now construct the spectrum of $\ham$. As discussed in \refsec{transfermatrix}, one can find simultaneous eigenstates of $\ham$ and $\Phi_{y}$. After substituting \refeqq{eq:czeta} into \refeqq{eq:yflux1}, the latter is given by
\begin{equation}
\label{eq:fluxzeta}
\Phi_{y} = -\frac{L_{x}}{2} + \sum_{k\in\mathbb{K}_{p}}\zeta_{k}^{\dagger}\zeta_{k}
\end{equation}
in terms of $\zeta_{k}$ fermions, which counts the number of occupied states relative to half filling [the number of available $k$-states is $L_{x}$ by \refeqqand{eq:qevenflux}{eq:qoddflux}].\footnote{$\Phi_{y}$ does not contain $C_{L_{x}+1}$ and so does not depend on $p$; either \(p\) gives the same result.}

The occupation-number states of the \(\zeta_k\) fermions with $k\in\mathbb{K}_{p}$ form a complete set of mutual eigenstates of $\ham_{p}$ and $\Phi_{y}$. From \refeqq{eq:hamprojected}, the complete set of eigenstates of \(\ham\) is given by the union of all eigenstates of $\ham_{0}$ that have even $\Phi_{y}$ eigenvalue and all eigenstates of $\ham_{1}$ that have odd $\Phi_{y}$ eigenvalue. We will denote $\lvert \Phi_{y} \rangle_{n}$ as the $n$th excited eigenstate of $\ham$ with vertical flux $\Phi_{y}$, and $E_{n}(\Phi_{y})$ as its eigenenergy. The spectrum of the two-row transfer matrix $W$ follows from that of $\ham$ through \refeqq{eq:ham}: $\lvert \Phi_{y} \rangle_{n}$ is also an eigenstate of $W$, but with eigenvalue $\ee^{-2E_{n}(\Phi_{y})}$.

\begin{figure*}
\begin{center}
\includegraphics[width=0.8\textwidth]{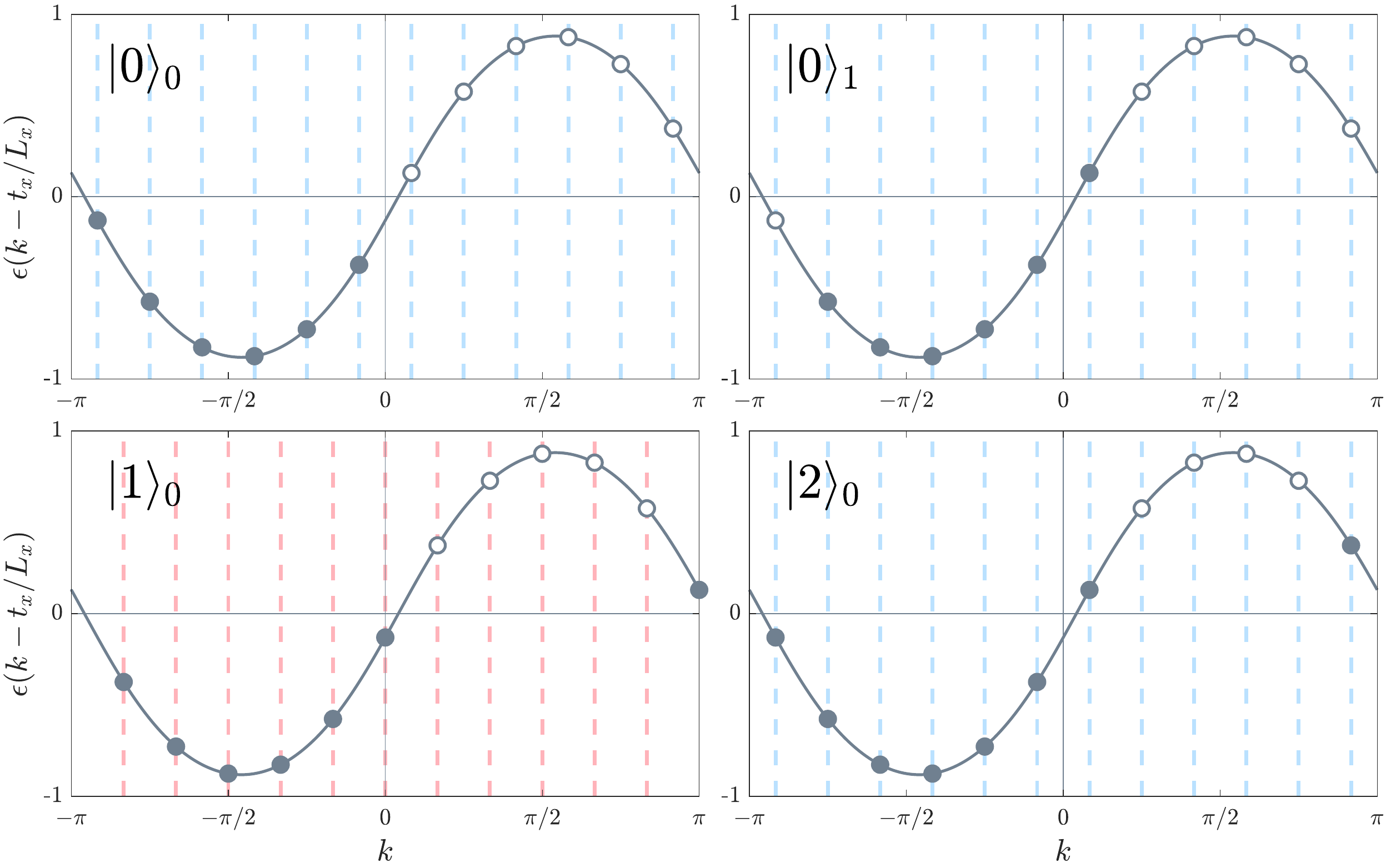}
\caption{Simultaneous eigenstates of the Hamiltonian $\ham$, given by \refeqqand{eq:hamprojected}{eq:ham2}, and the vertical flux $\Phi_{y}$ of \refeqq{eq:fluxzeta}, for $L_{x} = 12$, $\alpha = 1$ and $t_{x} = \pi/2$: The $n$th excited eigenstate with vertical flux $\Phi_{y}$ is denoted by $\lvert \Phi_{y}\rangle_{n}$, while filled and empty circles represent filled and empty $\zeta_{k}$ orbitals, respectively. Top-left panel: Ground state $\lvert 0 \rangle_{0}$, where $k$-states, given by \refeqq{eq:qevenflux} for $\Phi_{y}$ even (dashed blue lines), are all occupied for $\epsilon (k- t_{x}/L_{x}) < 0$. Top-right panel: First excited state in the $\Phi_{y} = 0$ sector $\lvert 0 \rangle_{1}$, obtained by adding a particle-hole excitation to $\lvert 0\rangle_{0}$. Bottom-left panel: Lowest-energy state in the $\Phi_{y} = 1$ sector $\lvert 1 \rangle_{0}$, where $k$-states, given by \refeqq{eq:qoddflux} for $\Phi_{y}$ odd (dashed red lines), are occupied for $-\pi \leq k \leq 0$. Bottom-right panel: Lowest-energy state in the $\Phi_{y} = 2$ sector $\lvert 2 \rangle_{0}$, obtained by adding two particles to $\lvert 0\rangle_{0}$.}
\label{fig:spectrum}
\end{center}
\end{figure*}

As illustrated in \reffig{fig:spectrum} (top-left panel), the ground-state is half filled and thus denoted by $\lvert 0 \rangle_{0}$. Formally, it is defined by
\begin{equation}
\begin{split}
&\zeta_{k}\lvert 0 \rangle_{0} = 0 \qquad \text{for} \qquad 0 < k < \pi\\
&\zeta_{k}^{\dagger}\lvert 0 \rangle_{0} = 0 \qquad \text{for} \qquad -\pi < k < 0 \punc ,
\end{split}
\end{equation}
where $k\in \mathbb{K}_{0}$, and has energy
\begin{equation}
\label{eq:groundstateenergy}
E_{0}(0) = \sum_{\substack{k\in\mathbb{K}_{0} \\ k < 0}}\epsilon(k-t_{x}/L_{x}) \punc .
\end{equation}
\reffig{fig:spectrum} also illustrates some eigenstates with higher energy.

To calculate the ground-state energy $E_{0}(0)$, in the limit $L_{x} \rightarrow \infty$ and including $O(1/L_{x})$ corrections, we rewrite the sum in \refeqq{eq:groundstateenergy} as an integral using the Euler--Maclaurin formula
\begin{multline}
\label{eq:EM}
\sum_{i=0}^{n}f(a + i\delta) = \frac{1}{\delta}\int_{a}^{a+n\delta}f(\phi) \, d\phi + {} \\ \frac{1}{2} [f(a) + f(a + n\delta)] + \frac{\delta}{12}[f'(a + n\delta) - f'(a)] + O(\delta^{3}) \punc ,
\end{multline}
with $a = -(L_{x}-1)\frac{\pi}{L_{x}}$, $\delta = \frac{2\pi}{L_{x}}$ and $n = \frac{L_{x}}{2} - 1$. The integral can be performed by extending the range of integration to $[-\pi, 0]$ and expanding $\epsilon(k-t_{x}/L_{x})$ as a power series in $1/L_{x}$. The leading term is then
\begin{equation}
\frac{L_{x}}{2\pi}\int_{-\pi}^{0} dk \, \sinh^{-1}(\alpha\sin k) = \frac{\ii L_{x}\chi_{2}(\ii \alpha)}{\pi} \punc ,
\end{equation}
where $\chi_{2}(z) = \frac{1}{2}[\operatorname{Li}_2(z)-\operatorname{Li}_2(-z)]$ is the Legendre chi function [in particular, $\chi_{2}(\ii) = \ii G$, where
\begin{equation}
G = \sum_{n=0}^{\infty}\frac{(-1)^{n}}{(2n+1)^{2}}
\end{equation}
is Catalan's constant]. The $O(L_{x}^{0})$ term vanishes, while the $O(1/L_{x})$ term is $t_{x}^{2}\alpha/2\pi L_{x}$.

The correction terms
\begin{equation}
-\frac{1}{\delta}\left[\int_{-\pi}^{a}  f(\phi)\, d\phi + \int_{a+n\delta}^{0} f(\phi) \, d\phi \right] \punc ,
\end{equation}
which arise when extending the integration bounds, as well as the remaining terms in \refeqq{eq:EM}, can be calculated using the Taylor expansion
\begin{equation}
\label{eq:dispersionseries}
\epsilon(k) = v\sub{F} k + O(k^3), \qquad \lvert k \rvert \ll 1 \punc ,
\end{equation}
where \(v\sub{F}=\alpha\) is the Fermi velocity. The final result is
\begin{equation}
\label{eq:e00}
E_{0}(0) = \frac{\ii L_{x}\chi_{2}(\ii \alpha)}{\pi} - \frac{\pi\alpha}{6L_{x}} + \frac{t_{x}^{2}\alpha}{2\pi L_{x}} + O\left(\frac{1}{L_{x}^3}\right) \punc ,
\end{equation}
and a similar calculation for the lowest-energy state in the $\Phi_{y} = 1$ sector gives
\begin{equation}
\label{eq:e01}
E_{0}(1) = E_{0}(0) + \frac{\pi\alpha}{2L_{x}} + O\left(\frac{1}{L_{x}^3}\right)\punc .
\end{equation}

Note that the \(t_x\) dependence of \(E_0(0)\) is the standard result \cite{Falomir1990} for the \(O(L_x^{-1})\) correction to the ground-state energy of fermions with a twist \(t_x\) in their boundary conditions (see \reffootnote{FootnoteTwistedBC}). Relating this to the (effective) central charge \(c\) \cite{Blote1986}, we have
\beq
-\frac{\pi c \alpha}{6L_x} = - \frac{\pi\alpha}{6L_{x}} + \frac{t_{x}^{2}\alpha}{2\pi L_{x}}\punc,
\eeq
and so
\beq
c = 1 - \frac{3t_x^2}{\pi^2}\punc.
\eeq
In particular, for \(t_x = 0\), this gives the expected result of \(c=1\) for a theory containing a single complex fermion. We note, however, that the value of the central charge in the dimer model is controversial, with other arguments suggesting instead \(c=-2\) when \(t_x=0\) \cite{MorinDuchesne2016}.

\section{Partition function}
\label{partitionfunction}

In this section, we write down the partition function \(Z(\bm{t})\) using \refeqq{eq:Ztrace} and eigenvalues of the two-row transfer matrix, before taking the thermodynamic limit.

By \refeqqand{eq:wprojectors}{eq:wphp}, one can split \(Z(\bm{t})\) into contributions from each parity sector, giving
\beq[eq:Zparitysectors]
Z(\bm{t}) = \Tr \left(\sum_{p} \ee^{\ii t_y \Phi_y}\ee^{-L_y \ham_p} \Pi_p\right)\punc.
\eeq
The projector \(\Pi_p\) can be expanded using \refeqqand{eq:definePi}{eq:fluxzeta} as
\beq
\Pi_p = \frac{1}{2}\sum_{\sigma=\pm} \sigma^p \exp\left[-\ii \pi \delta_{\sigma,-}\left(-\frac{L_{x}}{2} + \sum_{k\in\mathbb{K}_{p}}\zeta_{k}^{\dagger}\zeta_{k}\right)\right]\punc,
\eeq
and hence
\beq[eq:eePi]
\ee^{\ii t_y \Phi_y}\ee^{-L_y \ham_p} \Pi_p = \frac{1}{2}\sum_{\sigma=\pm} \sigma^p \ee^{-L_y \tilde\ham_{p,\sigma}}\punc,
\eeq
where
\beq[eq:tildeham]
\tilde\ham_{p,\sigma} = \frac{\ii L_x}{2L_y}(t_y - \pi \delta_{\sigma,-}) + \sum_{k \in \mathbb{K}_{p}} \tilde{\epsilon}_{\sigma}(k)\zeta_k^\dagger \zeta_k \punc ,
\eeq
with
\beq
\tilde{\epsilon}_{\sigma}(k) = \epsilon(k - t_x/L_x) - \frac{\ii}{L_y}( t_y - \pi\delta_{\sigma,-})\punc.
\eeq
The partition function, \refeqq{eq:Zparitysectors}, can therefore be written as
\beq[eq:Zstart]
Z(\bm{t}) = \frac{1}{2} \sum_{p,\sigma}\sigma^p Z_{p,\sigma}\punc,
\eeq
where \(Z_{p,\sigma} = \Tr \ee^{-L_y \tilde\ham_{p,\sigma}}\). Because the trace of an operator is equivalent to the sum of its eigenvalues, one has
\begin{equation}
\label{eq:Zsym}
Z_{p,\pm} = (\pm \ee^{\ii t_{y}})^{-L_{x}/2}\prod_{k\in\mathbb{K}_{p}}\left[1 \pm \ee^{-L_{y}\epsilon(k-t_{x}/L_{x})}\ee^{\ii t_{y}}\right] \punc,
\end{equation}
which reduces to Lieb's partition function for $\bm{t}=\bm{0}$ [see \refcite{Lieb1967}, Eq.~(3.14)].

We now take the thermodynamic limit, retaining leading-order corrections to the free-energy density. To do so for $Z_{0,\pm}$, we factor out $\pm \ee^{-L_{y}\epsilon(k-t_{x}/L_{x})}\ee^{\ii t_{y}}$ for all terms in the product with $k<0$ and restrict the product to $0<k\leq \pi/2$, which gives
\begin{equation}
\begin{split}
\label{eq:Zprod}
&Z_{0,\pm} = \ee^{-L_{y}E_{0}(0)}\times {} \\ &\Bigg\{\prod_{n=1}^{\ceil*{L_{x}/4}} \left[1\pm \ee^{-L_{y}\epsilon(k-t_{x}/L_{x})}\ee^{\ii t_{y}}\right]\left[1\pm \ee^{-L_{y}\epsilon(k+t_{x}/L_{x})}\ee^{-\ii t_{y}}\right]\Bigg\}\times {} \\
&\Bigg\{\prod_{n=1}^{\floor*{L_{x}/4}} \left[1\pm \ee^{-L_{y}\epsilon(k-t_{x}/L_{x})}\ee^{-\ii t_{y}}\right]\left[1\pm \ee^{-L_{y}\epsilon(k+t_{x}/L_{x})}\ee^{\ii t_{y}}\right]\Bigg\} \punc ,
\end{split}
\end{equation}
where $k = (2n-1)\frac{\pi}{L_{x}}$ by \refeqq{eq:qevenflux}.

In the limit $L_{x},L_{y}\rightarrow \infty$, we can replace $\epsilon(k\pm t_{x}/L_{x})$ by its leading-order dependence $\alpha(k\pm t_{x}/L_{x})$ [see \refeqq{eq:dispersionseries}], since the next-order terms will eventually be of order $L_{y}/L_{x}^{3}$. Hence, \refeqq{eq:Zprod} becomes
\begin{multline}
Z_{0,\pm} = \ee^{-L_{y}E_{0}(0)}\prod_{n=1}^{\infty}(1 \pm yq^{n-1/2})(1 \pm y^{-1}q^{n-1/2})\times {} \\ (1 \pm y^{*}q^{n-1/2})(1 \pm y^{*-1}q^{n-1/2}) \punc ,
\end{multline}
where $y = \ee^{\rho t_{x}}\ee^{\ii t_{y}}$, $q = \ee^{-2\pi\rho}$ and $\rho = \alpha L_{y}/L_{x}$. This can be expressed in terms of Jacobi theta functions using the first equality of \refeqqand{eq:t3}{eq:t4}:
\begin{equation}
Z_{0,+} = \ee^{-L_{y}E_{0}(0)}q^{1/12}\frac{\theta_{3}(y\vert q)\theta_{3}(y^*\vert q)}{\eta^{2}(q)} \punc ,
\end{equation}
where $\eta(q)$ is the Dedekind eta function defined in \refeqq{eq:dedekind}, and the same for $Z_{0,-}$ but with $\theta_{3} \rightarrow \theta_{4}$. An analogous calculation for $Z_{1,\pm}$ yields
\begin{equation}
Z_{1,+} = \ee^{-L_{y}E_{0}(1)}q^{-1/6}\frac{\theta_{2}(y\vert q)\theta_{2}(y^*\vert q)}{\eta^{2}(q)} \punc ,
\end{equation}
with $\theta_{2} \rightarrow \theta_{1}$ for $Z_{1,-}$.

Combining the results for $Z_{p,\sigma}$ with \refeqqand{eq:e00}{eq:e01}, \refeqq{eq:Zstart} becomes
\begin{multline}
\label{eq:ztorustu}
Z(\bm{t}) = \exp{\left[\frac{-\ii L_{x}L_{y}\chi_{2}(\ii \alpha)}{\pi}\right]}\exp{\left(-\frac{\rho t_{x}^{2}}{2\pi}\right)} \times {} \\ \frac{\sum_{i=1}^{4}\theta_{i}(y\vert q)\theta_{i}(y^{*}\vert q)}{2\eta^{2}(q)} \punc ,
\end{multline}
which is consistent with Eq.~(8.41) of \refcite{Rasmussen2012} when $t_{x}=0$. When $\bm{t}=\bm{0}$, $\theta_{1}(1\vert q) = 0$ and the partition function is
\begin{equation}
\label{eq:ztorus}
Z(\bm{0}) = \exp{\left[\frac{-\ii L_{x}L_{y}\chi_{2}(\ii \alpha)}{\pi}\right]}\frac{\sum_{i=2}^{4}\theta_{i}^{2}(1\vert q)}{2\eta^{2}(q)} \punc ,
\end{equation}
in agreement with \refcite{Ferdinand1967}.

The first factor in \refeqq{eq:ztorustu} grows exponentially with system volume, and represents the weight of dimer configurations in the bulk, i.e., it specifies the bulk free-energy density \cite{Rasmussen2012}
\begin{align}
f\sub{bulk} &= -\lim_{L_{x},L_{y}\rightarrow \infty} \frac{1}{L_{x}L_{y}}\log Z(\bm{t}) \\
&= \frac{\ii\chi_{2}(\ii \alpha)}{\pi} \punc .
\end{align}
As one might expect, $f\sub{bulk}$ does not depend on the choice of boundary conditions, although we note that this is not true in the case of the honeycomb lattice \cite{Elser1984}. 

The remaining terms in $Z(\bm{t})$, which give leading finite-size corrections to the free-energy density, are boundary dependent and, in the case of PBCs, encode information about topological flux sectors (see subsection below). Previously, these terms have also been evaluated (for \(\bm{t} = \bm{0}\)) with closed \cite{Ferdinand1967} and cylindrical \cite{McCoy2014} boundaries, as well as embeddings on the M\"{o}bius strip and Klein bottle \cite{Lu1999}. In general, one obtains terms in the free energy proportional to the edge of the system [e.g., $2(L_{x}+L_{y})$ for closed boundaries] and of order $L_{y}/L_{x}$. However, with PBCs (i.e., a torus) the edge is zero and we only observe the latter.

Using the modular identities given in \refapp{jacobithetafunctions}, one can confirm that the partition function \(Z(\bm{t})\) behaves as expected under \(90^\circ\) rotations, in spite of the asymmetry between \(x\) and \(y\) in the transfer-matrix method. Such a rotation takes \((t_x,t_y)\rightarrow(-t_y,t_x)\), while swapping \(L_x \leftrightarrow L_y\) and making the replacement \(\alpha^{N_x} \rightarrow \alpha^{N_y} = \alpha^{\frac{1}{2}L_xL_y - N_x}\) in the definition of the partition function, \refeqq{eq:partitionfunction}. We therefore expect \(Z(t)\) to be multiplied by \(\alpha^{\frac{1}{2}L_xL_y}\) while \(\rho = \alpha L_y/L_x\) becomes \(1/\rho\).

From \refeqqand{eq:modularidentity}{eq:modularidentityrho}, we find
\begin{multline}
\exp{\left(-\frac{\rho t_{x}^{2}}{2\pi}\right)} \frac{\sum_{i=1}^{4}\theta_{i}(y\vert q)\theta_{i}(y^{*}\vert q)}{2\eta^{2}(q)}\\=\exp{\left(-\frac{ t_{y}^{2}}{2\pi\rho}\right)} \frac{\sum_{i=1}^{4}\theta_{i}(y'\vert q')\theta_{i}({y'}^{*}\vert q')}{2\eta^{2}(q')}\punc,
\end{multline}
where \(y' = \ee^{-t_y/\rho}\ee^{\ii t_x}\) and \(q' = \ee^{-2\pi/\rho}\) correspond to \(y\) and \(q\) under rotation. Since \(\chi_2(\ii/\alpha) = \chi_2(\ii\alpha) - \frac{\ii \pi}{2}\log \alpha\) \cite[Sec.~25.12]{NIST:DLMF}, the remaining (bulk) factor in \refeqq{eq:ztorustu} is replaced by
\beq
\exp{\left[\frac{-\ii L_{y}L_{x}\chi_{2}(\ii/\alpha)}{\pi}\right]} = \alpha^{\frac{1}{2}L_xL_y}\exp{\left[\frac{-\ii L_{x}L_{y}\chi_{2}(\ii\alpha)}{\pi}\right]}\punc,
\eeq
giving the expected transformation of \(Z(\bm{t})\).

\subsection*{Flux sectors}

We now show how the partition function, \refeqq{eq:ztorus}, divides into topological sectors labeled by the flux. By construction, $Z(\bm{t})$ is periodic in $t_{\mu}$ (with period $2\pi$), so can be expressed as a Fourier series
\begin{equation}
\label{eq:zfs}
Z(\bm{t}) = \sum_{\bm{\Phi}}\tilde{Z}_{\bm{\Phi}}\ee^{\ii \bm{t}\cdot\bm{\Phi}} \punc .
\end{equation}
Comparison of \refeqqand{eq:partitionfunction}{eq:zfs} implies \begin{equation}
\label{eq:zfs2}
\tilde{Z}_{\bm{\Phi}} = \sum_{c\in\goC_0(\bm{\Phi})} \alpha^{N_{x}} \punc ,
\end{equation}
where the set $\goC_0(\bm{\Phi})$ contains all close-packed dimer configurations with flux $\bm{\Phi}$. In other words, the Fourier coefficient $\tilde{Z}_{\bm{\Phi}}$ can be interpreted as the partial partition function, or total weight, of flux sector $\bm{\Phi}$.

To calculate $\tilde{Z}_{\bm{\Phi}}$, we use the second equality of \refeqqs{eq:t1}{eq:t4} to rewrite \refeqq{eq:ztorustu} as \cite{Boutillier2009}
\begin{equation}
Z(\bm{t}) = \ee^{-L_{x}L_{y}f\sub{bulk}}\frac{\sum_{m \in \mathbb{Z}}\ee^{-\rho(t_{x} - 2\pi m)^2/2\pi}\sum_{n\in\mathbb{Z}}\ee^{\ii nt_{y}}\ee^{-\pi\rho n^{2}/2}}{\eta^{2}(q)}
\end{equation}
(the periodicity in $t_{x}$ is now apparent). The sum over $m$ can be written in the same form as the sum over $n$ through the Poisson summation formula, giving
\begin{equation}
\label{eq:zsums}
Z(\bm{t}) = \ee^{-L_{x}L_{y}f\sub{bulk}}\frac{\sum_{m \in \mathbb{Z}}\ee^{\ii mt_{x}}\ee^{-\pi m^{2}/2\rho}\sum_{n\in\mathbb{Z}}\ee^{\ii nt_{y}}\ee^{-\pi\rho n^{2}/2}}{\sqrt{2\rho}\eta^{2}(q)} \punc ,
\end{equation}
which allows us to read off from \refeqqand{eq:zfs}{eq:zsums}
\begin{equation}
\tilde{Z}_{\bm{\Phi}} = \ee^{-L_{x}L_{y}f\sub{bulk}}\frac{\ee^{-\pi (\Phi_{x}^{2}/\rho + \rho\Phi_{y}^{2})/2}}{\sqrt{2\rho}\eta^{2}(q)} \punc .
\end{equation}
This result has previously been obtained for the honeycomb-lattice dimer model using Pfaffian methods \cite{Boutillier2009}, while \refcite{Rasmussen2012} has used the transfer matrix to calculate the partial partition function of flux sector $\Phi_{y}$, equivalent to $\sum_{\Phi_{x}}\tilde{Z}_{\bm{\Phi}}$ [see their Eqs.~(8.19) and (8.36)].

Knowledge of $\tilde{Z}_{\bm{\Phi}}$ can be used to calculate flux moments.
The probability of flux $\bm{\Phi}$ is given by
\begin{align}
\label{eq:fluxprob}
P(\bm{\Phi}) &= \frac{\tilde{Z}_{\bm{\Phi}}}{\sum_{\bm{\Phi}}\tilde{Z}_{\bm{\Phi}}} \\
& = \frac{\ee^{-\pi (\Phi_{x}^{2}/\rho + \rho\Phi_{y}^{2})/2}}{\sum_{m,n\in\mathbb{Z}}\ee^{-\pi (m^{2}/\rho + \rho n^{2})/2}} \punc ,
\end{align}
which implies that $\Phi_{x}$ and $\Phi_{y}$ are independent variables. This form is known from effective field theories \cite{Alet2006b,Tang2011}. The mean flux vanishes by symmetry, while the mean-square flux is given by
\begin{equation}
\label{eq:phix2}
\langle \Phi_{x}^{2} \rangle = \frac{\sum_{n\in\mathbb{Z}}n^{2}\ee^{-\pi n^{2}/2\rho}}{\sum_{n\in\mathbb{Z}}\ee^{-\pi n^{2}/2\rho}} \punc ,
\end{equation}
and the same for $\Phi_{y}$ but with $\rho \rightarrow 1/\rho$.

\section{Expectation values}
\label{expectationvalues}

In this section, we compute various expectation values in the thermodynamic limit, using the spectrum of the two-row transfer matrix.

We use \refeqq{eq:correlatortrace}, and restrict to operators $O$ that conserve parity of \(\Phi_y\), i.e., \([O, (-1)^{\Phi_y}]=0\). From \refeqq{eq:yflux1}, this includes any product of an even number of $C_{j}$ fermions, and hence any operator constructed from \(d_{j,x}\) and \(d_{j,y}\) [see \refeqqand{eq:dxc}{eq:dyc}]. It also allows us to calculate the monomer distribution function, as we show in \refsec{monomerdistributionfunction}. With this restriction, and because $(-1)^{\Phi_{y}}$ commutes with any quadratic form in fermions, \(O(l) = U(l)^{-1} O U(l)\) can be written as
\begin{align}
\label{eq:olp}
O(l) &= O(l)\sum_{p}\Pi_{p} \\
\label{eq:olp2}
&= \sum_{p}O(l)_{p}\Pi_{p} \punc ,
\end{align}
where
\begin{equation}
\label{eq:Qlp}
O(l)_{p} = U_p(l)^{-1} O_{p} U_p(l)\punc ,
\end{equation}
and \(U_p(l)\) is given by \refeqq{eq:Ul} but with \(V\) replaced by \(V_p\).

As for the partition function, the trace in \refeqq{eq:correlatortrace} can be split into parity sectors by inserting Eqs.~(\ref{eq:wprojectors}), (\ref{eq:wphp}) and (\ref{eq:olp2}), which yields
\begin{equation}
\langle O'(l')O(l) \rangle =\\ \frac{1}{Z(\bm{t})}\sum_p \Tr\left[\ee^{\ii t_{y}\Phi_{y}}\ee^{-L_y \ham_p} \Pi_p O'(l')_p O(l)_p\right] \punc ,
\end{equation}
where we have used $[V_{p},\Pi_{p}]=0$ and assumed $[O_{p},\Pi_{p}]=0$ (it is always possible to choose $O_{p}$ in this way). By \refeqq{eq:eePi}, this can be rewritten as
\begin{equation}
\label{eq:Qexpval}
\langle O'(l')O(l) \rangle = \frac{\sum_{p,\sigma} \sigma^p Z_{p,\sigma} \langle O'(l') O(l) \rangle_{p,\sigma}}
{\sum_{p,\sigma}\sigma^p Z_{p,\sigma}} \punc ,
\end{equation}
where, assuming \(Z_{p,\sigma}\neq 0\),
\beq[eq:Qexpvalpsigma]
\langle O'(l') O(l) \rangle_{p,\sigma} = \frac{1}{Z_{p,\sigma}}\Tr \left[ \ee^{-L_y \tilde\ham_{p,\sigma}} O'(l')_p O(l)_p\right]\punc.
\eeq
Expectation values are therefore given by an average over the four \((p,\sigma)\) sectors, each weighted by \(Z_{p,\sigma}\).

\subsection{Two-point correlation functions of $C_{j}$ fermions}

For an operator \(O\) given by a product of \(C_j\) fermions, the corresponding time-evolved operator \(O(l)_p\) can also be expressed as a product of \(C_j(l)_p\), with the same \(p\) for each. For example, when $O = d_{j,y}$ one has
\begin{align}
d_{j,y}(l)_{p} &= U_p(l)^{-1} C_j^\dagger C_j U_p(l)\\
&= U_p(l)^{-1} C_j^\dagger U_p(l) U_p(l)^{-1} C_j U_p(l)\\
&= C_j^\dagger(l)_{p} C_j(l)_{p}\punc .
\end{align}
Here, \(C_j(l)_p\) is defined by extending the definition in \refeqq{eq:Qlp} to \(C_j\), even though it does not conserve parity and so does not obey \refeqq{eq:olp2}.

An expectation value \(\langle O'(l') O(l) \rangle_{p,\sigma}\) can then be expressed in terms of a product of an even number of $C_{j}(l)$ operators. Because this is a time-ordered product and \(\tilde\ham_{p,\sigma}\) is a free-fermion Hamiltonian, Wick's theorem \cite{Abrikosov2012} applies, which allows us to write $\langle O'(l')O(l)\rangle_{p,\sigma}$ as a sum over products of two-point $C_{j}(l)$ correlators in each $(p,\sigma)$ sector. [We similarly extend the definition \refeqq{eq:Qexpvalpsigma} to include \(O=C_j\), even though \refeqq{eq:Qexpval} is not valid in this case.] We calculate these two-point correlators in this section.

To do so, we first use \refeqqand{eq:Ul}{eq:Qlp} to derive an expression for $C_{j}(l)_p$ in terms of $\zeta_{k}$ fermions. For $l$ even, \refeqq{eq:ham2} implies
\beq[eq:eveny]
W_p^{-1}\zeta_k W_p = \ee^{-2\epsilon(k-t_x/L_x)}\zeta_k \punc ,
\eeq
which can be used in \refeqq{eq:czeta} to give
\begin{multline}
\label{eq:Cjyeven}
C_{j}(l)_p =
\sqrt{\frac{2}{L_{x}}}\ee^{-\ii \pi/4}\sum_{k\in\mathbb{K}_{p}}\ee^{\ii kj} \times{}\\
\begin{cases}\cos\theta_{k+t_x/L_x}\ee^{-l\epsilon(k+t_x/L_x)}\zeta_{-k}^\dagger& \text{for \(j\) odd}\\
\cos\theta_{k-t_x/L_x}\ee^{-l\epsilon(k-t_x/L_x)}\zeta_k& \text{\phantom{for} \(j\) even.}
\end{cases}
\end{multline}
For $l$ odd, as well as \refeqq{eq:eveny} we additionally require the results
\beq[eq:onerowzeta]
\begin{aligned}
(V_p^{\dagger})^{-1}\zeta_{k}^*V_p^{\dagger} &= - \ee^{-\epsilon(k-t_x/L_x)}\zeta_{k-\pi}^{\dagger}\\
(V_p^{\dagger})^{-1}{\big(\zeta_{k}^\dagger\big)}^*V_p^{\dagger} &= - \ee^{\epsilon(k-t_x/L_x)}\zeta_{k-\pi}\punc,
\end{aligned}
\eeq
which can be derived from \refeqq{eq:tmjv}. This time we use these in the complex conjugate of \refeqq{eq:czeta}, to find
\begin{multline}
\label{eq:Cjyodd}
C_{j}(l)_p =
\sqrt{\frac{2}{L_{x}}}\ee^{\ii \pi/4}\sum_{k\in\mathbb{K}_{p}}\ee^{\ii kj} \times{}\\
\begin{cases}\phantom{-}\cos\theta_{k-t_x/L_x} \ee^{-l\epsilon(k - t_{x}/L_{x})}\zeta_{k}& \text{for \(j\) odd}\\
-\cos\theta_{k+t_x/L_x}\ee^{-l\epsilon(k+t_x/L_x)}\zeta_{-k}^{\dagger}& \text{\phantom{for} \(j\) even.}
\end{cases}
\end{multline}
Finally, by combining \refeqqand{eq:Cjyeven}{eq:Cjyodd}, we have
\begin{multline}
\label{eq:Cjy}
C_{j}(l)_p =
\sqrt{\frac{2}{L_{x}}}\ee^{-\ii(-1)^l \pi/4}\sum_{k\in\mathbb{K}_{p}}\ee^{\ii k j} \times{}\\
\begin{cases}(-1)^l\cos\theta_{k+t_x/L_x} \ee^{-l\epsilon(k + t_{x}/L_{x})}\zeta_{-k}^\dagger& \text{for \(j+l\) odd}\\
\phantom{(-1)^l}\cos\theta_{k-t_x/L_x}\ee^{-l\epsilon(k-t_x/L_x)}\zeta_k& \text{\phantom{for} \(j+l\) even,}
\end{cases}
\end{multline}
for all $l$.

Since \(\tilde\ham_{p,\sigma}\), defined in \refeqq{eq:tildeham}, is a free-fermion Hamiltonian with dispersion \(\tilde\epsilon_{\sigma}\), and \refeqq{eq:Qexpvalpsigma} describes a thermal distribution with effective temperature \(1/L_y\), the two-point correlation functions of the \(\zeta_{k}\) fermions are given by
\beq
\label{eq:zetacorrelators}
\begin{aligned}
\langle \zeta_{k}\zeta_{k'} \rangle_{p,\sigma} &= \langle \zeta_{k}^{\dagger}\zeta_{k'}^{\dagger} \rangle = 0\\
\langle \zeta_{k}^{\dagger}\zeta_{k'} \rangle_{p,\sigma} &= \delta_{kk'}\nF\bm(L_y \tilde\epsilon_{\sigma}(k)\bm)\\
\langle \zeta_{k}\zeta_{k'}^{\dagger} \rangle_{p,\sigma} &= \delta_{kk'}\nF\bm(-L_y \tilde\epsilon_{\sigma}(k)\bm)\punc ,
\end{aligned}
\eeq
where \(\nF(z) = (\ee^z + 1)^{-1}\) is the Fermi--Dirac distribution function.

{
\allowdisplaybreaks
Hence, denoting $\Rv = (X,Y)$, the $C_{j}(l)$ correlators are
\begin{widetext}
\begin{align}
\label{eq:cc}
\langle C_{j+X}(l+Y)C_{j}(l) \rangle_{p,\sigma} &=
\begin{cases}
-\ee^{\ii\varphi(l,Y)} \Gamma_{p,\sigma}(\Rv, -\bm t) & \text{for \(X+Y\) odd, $j+l$ odd} \\
-\ee^{-\ii\varphi(l,Y)}\Gamma_{p,\sigma}(\Rv, \bm t) & \text{\phantom{for} \(X+Y\) odd, $j+l$ even}\\
0 & \text{\phantom{for} \(X+Y\) even}
\end{cases} \\
\label{eq:cdcd}
\langle C_{j+X}^{\dagger}(l+Y)C_{j}^{\dagger}(l) \rangle_{p,\sigma} &=
\begin{cases}
\ee^{-\ii\varphi(l,Y)} \Gamma_{p,\sigma}(\Rv, \bm t) & \text{for \(X+Y\) odd, $j+l$ odd} \\
\ee^{\ii\varphi(l,Y)}\Gamma_{p,\sigma}(\Rv, -\bm t) & \text{\phantom{for} \(X+Y\) odd, $j+l$ even}\\
0 & \text{\phantom{for} \(X+Y\) even}
\end{cases} \\
\label{eq:cdc}
\langle C_{j+X}^{\dagger}(l+Y)C_{j}(l) \rangle_{p,\sigma} &=  \begin{cases}
0 & \text{for \(X+Y\) odd} \\
\ee^{\ii\varphi(l,Y)} \left[\Delta_{p,\sigma}(\Rv, -\bm t) - \Gamma_{p,\sigma}(\Rv, -\bm t)\right] & \text{\phantom{for} \(X+Y\) even, $j+l$ odd,}\\
\ee^{-\ii\varphi(l,Y)} \left[\Delta_{p,\sigma}(\Rv, \bm t) - \Gamma_{p,\sigma}(\Rv, \bm t)\right] & \text{\phantom{for} \(X+Y\) even, $j+l$ even}
\end{cases} \\
\label{eq:ccd}
\langle C_{j+X}(l+Y)C_{j}^{\dagger}(l) \rangle_{p,\sigma} &=  \begin{cases}
0 & \text{for \(X+Y\) odd} \\
\ee^{-\ii\varphi(l,Y)} \left[\Delta_{p,\sigma}(\Rv, \bm t) + \Gamma_{p,\sigma}(\Rv, \bm t)\right] & \text{\phantom{for} \(X+Y\) even, $j+l$ odd,}\\
\ee^{\ii\varphi(l,Y)} \left[\Delta_{p,\sigma}(\Rv, -\bm t) + \Gamma_{p,\sigma}(\Rv, -\bm t)\right] & \text{\phantom{for} \(X+Y\) even, $j+l$ even}
\end{cases}
\end{align}
\end{widetext}
where
\begin{equation}
\varphi(l,Y) = \begin{cases}
(-1)^{l}\frac{\pi}{2} & \text{for $Y$ odd} \\
0 & \text{\phantom{for} $Y$ even,}
\end{cases}
\end{equation}
and
\begin{align}
\begin{split}
\Gamma_{p,\sigma}(\Rv, \bm t) &= \frac{1}{L_x}\sum_{k \in \mathbb{K}_p}\ee^{-\ii kX}\ee^{Y\epsilon(k-t_{x}/L_{x})}\nF\bm(L_y \tilde\epsilon_{\sigma}(k)\bm) \times {} \\
&\qquad\begin{cases}
\phantom{-}\ii\sin(2\theta_{k-t_{x}/L_{x}}) & \text{for $X+Y$ odd} \\
-\cos(2\theta_{k-t_{x}/L_{x}}) & \text{\phantom{for} $X+Y$ even}
\end{cases}
\end{split} \\
\Delta_{p,\sigma}(\Rv, \bm t) &= \frac{1}{L_x}\sum_{k \in \mathbb{K}_p}\ee^{-\ii kX}\ee^{Y\epsilon(k-t_{x}/L_{x})}\nF\bm(L_y \tilde\epsilon_{\sigma}(k)\bm) \punc .
\end{align}
These results are exact, with the correct (anti)periodicity in the horizontal direction, and could be used to calculate expectation values for finite system sizes as a function of flux sector.
}

Instead, we take the thermodynamic limit \(L_x,L_y\rightarrow \infty\), keeping the ratio \(L_y/L_x\) and the separation \(\lvert \Rv \rvert\) finite. In this limit, \(\nF(z)\) can be replaced by a step function \(\vartheta(-\operatorname{Re} z)\) and the discrete \(k\) values become continuous, giving
\begin{align}
\label{eq:gammaint}
\begin{split}
\Gamma_{p,\sigma}(\Rv,\bm t) &\approx \Gamma(\Rv) = \int_{0}^{\pi} \frac{dk}{2\pi}\, \ee^{\ii kX}\ee^{-Y\epsilon(k)} \times {} \\
&\qquad\qquad\begin{cases}
\ii \sin(2\theta_{k}) & \text{for $X+Y$ odd} \\
\cos(2\theta_{k}) & \text{\phantom{for} $X+Y$ even}
\end{cases}
\end{split} \\
\label{eq:deltaint}
\Delta_{p,\sigma}(\Rv,\bm t) &\approx \Delta({\Rv}) = \int_{0}^{\pi} \frac{dk}{2\pi}\, \ee^{\ii kX}\ee^{-Y\epsilon(k)} \punc .
\end{align}
Some values of these integrals for small $\lvert \Rv\rvert$ are shown in \reftab{tab:gvalues}, expressed in terms of the quantities
\begin{equation}
\rho_{x} = \frac{\arctan{\alpha}}{\pi} \qquad
\rho_{y} = \frac{\arctan{(1/\alpha)}}{\pi} \punc ,
\end{equation}
which satisfy $\rho_{x} + \rho_{y} = \frac{1}{2}$. For large $\lvert \Rv\rvert$, the asymptotic behavior is obtained by integrating by parts repeatedly, treating the cases $Y\gg 1$ [where \refeqq{eq:dispersionseries} can be used] and $Y$ of order unity separately.

\renewcommand{\arraystretch}{2.5}
\begin{table}
\begin{center}
\begin{tabular}{ | P{3.8cm} |  P{3.8cm} |}
\hline
Integral & Value \\ \hline
$\Gamma(\bm{0})$ & $\rho_{x}$ \\ \hline
$\Gamma(1,0)$ & $-\dfrac{\rho_{x}}{\alpha}$ \\ \hline 
$\Gamma(0,1)$ & $\ii\rho_{y}$ \\ \hline
$\Gamma(2,0)$ & $ -\dfrac{1}{\pi\alpha} + \dfrac{\rho_{x}}{\alpha^{2}}$ \\ \hline
$\Gamma(1,2)$ & $-\dfrac{1}{\pi} + \alpha\rho_{y}$ \\ \hline
$\Gamma(2,1)$ & $-\dfrac{\ii}{\alpha^{2}}\left(\rho_{x} - \dfrac{\alpha}{\pi}\right)$ \\ \hline
$\Gamma(3,0)$ & $-\rho_{x}\left(\dfrac{1}{\alpha} + \dfrac{2}{\alpha^{3}}\right) + \dfrac{2}{\pi\alpha^{2}}$ \\ \hline
$\Gamma(0,3)$ & $\ii\left[\rho_{y}(1+2\alpha^{2}) - \dfrac{2\alpha}{\pi}\right]$ \\ \hline
$\Gamma(\lvert \bm{R} \rvert\gg 1)$, $X$ odd, $Y$ even & $-\dfrac{1}{\pi}\dfrac{X}{X^{2} + (\alpha Y)^{2}}$ \\ \hline
$\Gamma(\lvert \bm{R} \rvert\gg 1)$, $X$ even, $Y$ odd & $\dfrac{\ii}{\pi}\dfrac{\alpha Y}{X^{2} + (\alpha Y)^{2}}$  \\ \hline
$\Gamma(\lvert \bm{R} \rvert\gg 1)$, $X$ odd, $Y$ odd & $\dfrac{2 \ii \alpha}{\pi}\dfrac{X\alpha Y}{[X^{2} + (\alpha Y)^{2}]^{2}}$ \\ \hline
$\Gamma(\lvert \bm{R} \rvert\gg 1)$, $X$ even, $Y$ even & $-\dfrac{\alpha}{\pi}\dfrac{X^{2} - (\alpha Y)^{2}}{[X^{2} + (\alpha Y)^{2}]^{2}}$ \\ \hline
$\Delta(X \text{even},0)$ & $\frac{1}{2}\delta_{X,0}$ \\ \hline
$\Delta(\lvert\Rv\rvert \gg 1)$, $X$ odd & $\dfrac{\ii}{\pi}\dfrac{X}{X^{2} + (\alpha Y)^{2}}$ \\ \hline
$\Delta(\lvert\Rv\rvert \gg 1)$, $X$ even & $\dfrac{1}{\pi}\dfrac{\alpha Y}{X^{2} + (\alpha Y)^{2}}$ \\ 
\hline
\end{tabular}
\caption{Values of the integrals $\Gamma(\Rv)$ and $\Delta(\Rv)$, defined in \refeqqand{eq:gammaint}{eq:deltaint}, respectively, for small $\lvert \Rv \rvert$, as well as their asymptotic behavior for $\lvert \Rv \rvert \gg 1$. Values for $X < 0$ may be obtained using the relation $\Gamma(-X,Y) = (-1)^{X}\Gamma(\Rv)$ and the same for $\Delta(\Rv)$.}
\label{tab:gvalues}
\end{center}
\end{table}
\renewcommand{\arraystretch}{1}

These expressions are independent of \(p\) and \(\sigma\), i.e., all four \((p,\sigma)\) sectors make equal contributions in the thermodynamic limit. Hence, \refeqq{eq:Qexpval} is redundant to this order, and we simply have \(\langle O'(l')O(l) \rangle = \langle O'(l') O(l) \rangle_{0,+}\) for operators \(O\) that are products of an even number of $C_{j}$ fermions. We therefore drop the \((p,\sigma)\) indices from now on.

Furthermore, they are independent of \(\bm{t}\), whose leading-order dependence is \(O(L_x^{-1},L_y^{-1})\). This implies that expectation values are the same in any fixed flux sector in the thermodynamic limit (but note that that we have taken \(L_x,L_y\rightarrow\infty\), so this does not apply for $\bm{\Phi}\sim L_x,L_y$). To see this we rewrite \refeqq{eq:expectationvalue} as a sum over Fourier modes [cf. \refeqqand{eq:zfs}{eq:zfs2}]
\begin{equation}
\label{eq:O1}
\langle O \rangle = \frac{1}{Z(\bm{t})}\sum_{\bm{\Phi}}\langle O \rangle_{\bm{\Phi}}\tilde{Z}_{\bm{\Phi}}\ee^{\ii\bm{t}\cdot\bm{\Phi}} \punc ,
\end{equation}
where
\begin{equation}
\langle O \rangle_{\bm{\Phi}} = \frac{1}{\tilde{Z}_{\bm{\Phi}}}\sum_{c\in\goC_0(\bm{\Phi})}O\alpha^{N_{x}}
\end{equation}
is the expectation value of the observable $O$ in a fixed flux sector $\bm{\Phi}$. After multiplying both sides of \refeqq{eq:O1} by $Z(\bm{t})\ee^{-\ii\bm{t}\cdot\bm{\Phi}'}$ and integrating over $\bm{t}$, one finds that $\langle O \rangle_{\bm{\Phi}}=\langle O \rangle$ when the latter is independent of $\bm{t}$.

In subsequent sections we use \refeqqs{eq:cc}{eq:ccd} to calculate various observables in the dimer model in the thermodynamic limit. We expect our results to reproduce those of \refcite{Fisher1963} in this limit, since the choice of boundary conditions (PBCs versus closed) becomes irrelevant. We also note that asymptotic behavior of correlation functions can be predicted using effective field theories, although the results depend on phenomenological parameters known as the stiffnesses \cite{Tang2011}.

\subsection{Dimer occupation numbers}
\label{dimeroccupationnumber}

We first calculate the probability that a vertical or horizontal bond is occupied by a dimer, given by $\langle d_{j,y}(l) \rangle$ and $\langle d_{j,x}(l) \rangle$, respectively. (In the thermodynamic limit, there is no \(\bm{t}\) dependence, and so \(d_{j,x}^* = d_{j,x}\).)

Using \refeqqand{eq:dxc}{eq:dyc}, one finds
\begin{align}
\langle d_{j,x}(l) \rangle &=  -\alpha \Gamma(1,0)\\
\label{eq:drx}
&= \rho_{x} \punc ,
\end{align}
and
\begin{align}
\langle d_{j,y}(l) \rangle &= \Delta(\bm{0}) - \Gamma(\bm{0}) \\
\label{eq:dyvac}
&= \rho_{y}
\punc ,
\end{align}
consistent with Sec.~5 of \refcite{Fisher1963}. As required, each lattice site is touched by a dimer with probability unity, since $\langle d_{j,x}(l) \rangle + \langle d_{j,y}(l) \rangle = \frac{1}{2}$. In the isotropic case, $\alpha = 1$, one has $\langle d_{j,x}(l) \rangle = \langle d_{j,y}(l) \rangle = \frac{1}{4}$, whereas in the limit $\alpha \rightarrow 0$ ($\alpha \rightarrow \infty$) only vertical (horizontal) bonds are occupied.

\subsection{Dimer--dimer correlation functions}
\label{dimer--dimercorrelationfunctions}

Due to the close-packing constraint, the occupation of a given bond by a dimer is influenced by dimers far away. Hence, dimer--dimer correlations are non-trivial even in the absence of interactions. In this section, we show how they can be calculated by extending the above discussion to two-point correlators of $d_{j,x}$ and $d_{j,y}$.

\begin{figure}
\begin{center}
\includegraphics[width=0.9\columnwidth]{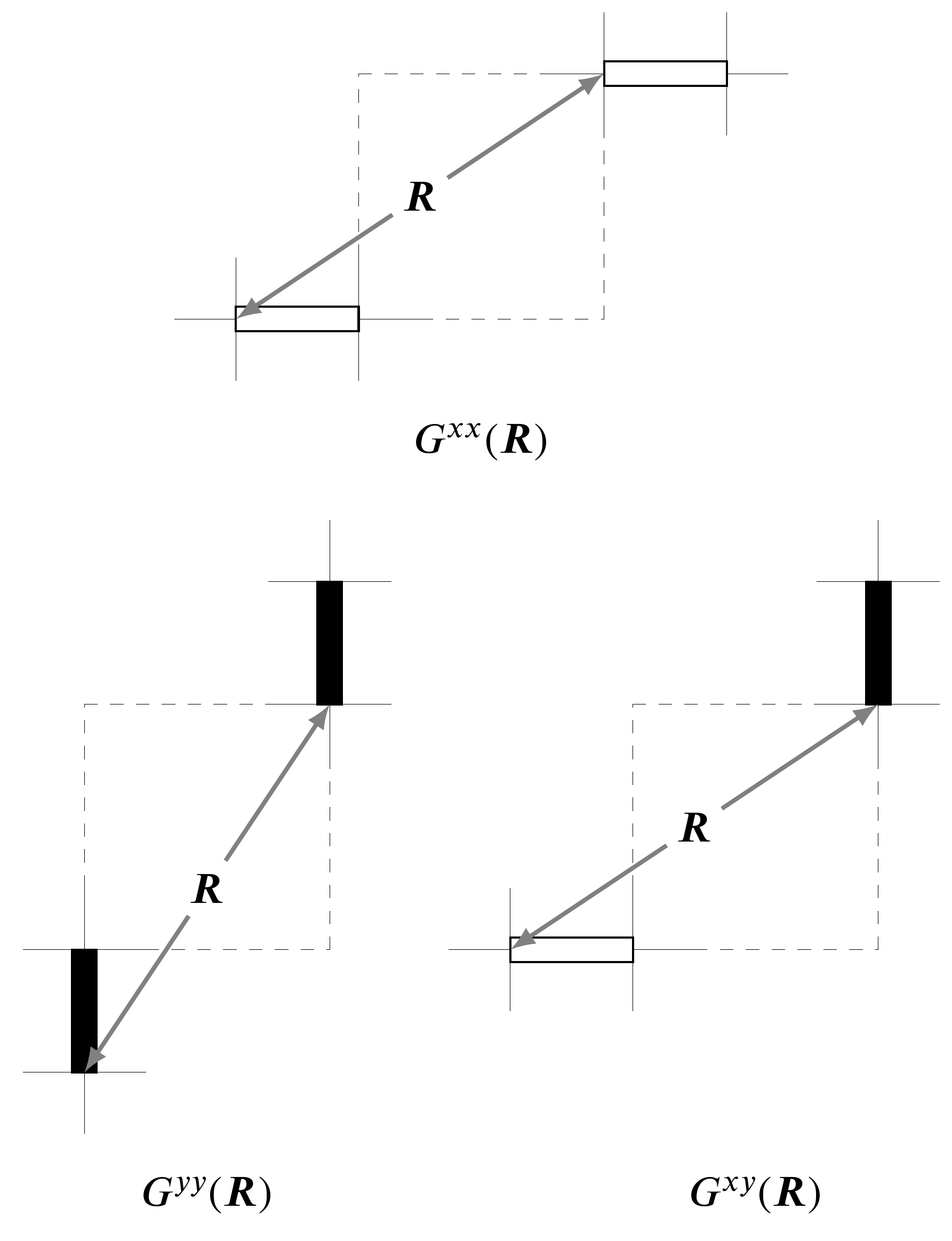}
\caption{Dimer--dimer correlation function between two horizontal dimers (top), two vertical dimers (bottom left), and a horizontal and vertical dimer (bottom right). In each case, the disconnected part of the correlator [i.e., the first term in Eqs.~(\ref{eq:gxx}), (\ref{eq:gyy}) and (\ref{eq:gxy})] is equal to the probability that the two bonds with separation $\Rv$ are both occupied.}
\label{fig:ddcf}
\end{center}
\end{figure}

The connected correlation function of two horizontal dimers with separation $\Rv$, illustrated in \reffig{fig:ddcf} (top), is given by (we assume $Y>0$ throughout this section)
\begin{multline}
\label{eq:gxx}
G^{xx}(\Rv) = \langle d_{j+X,x}(l+Y)d_{j,x}(l)\rangle - {} \\ \langle d_{j+X,x}(l+Y)\rangle\langle d_{j,x}(l)\rangle, \qquad \bm{R}\neq\bm{0}
\end{multline}
[for $\Rv=\bm{0}$ the first term vanishes due to $C_{j}^{2}(l)=0$; see \reffootnote{note:djx}]. Inserting \refeqq{eq:dxc} and using Wick's theorem \cite{Abrikosov2012} yields
\begin{multline}
\frac{G^{xx}(\Rv)}{\alpha^{2}} = \langle C_{j+X+1}(l+Y)C_{j}(l)\rangle\langle C_{j+X}(l+Y)C_{j+1}(l)\rangle - {} \\ \langle C_{j+X+1}(l+Y)C_{j+1}(l)\rangle\langle C_{j+X}(l+Y)C_{j}(l)\rangle \punc ,
\end{multline}
hence, by \refeqq{eq:cc},
\begin{equation}
\frac{G^{xx}(\Rv)}{\alpha^{2}} =
\begin{cases}
-\Gamma(\Rv)^{2} & \text{for \(X+Y\) odd} \\
\Gamma(X-1,Y)\Gamma(X+1,Y) & \text{\phantom{for} \(X+Y\) even.}
\end{cases}
\end{equation}
From \reftab{tab:gvalues}, some values for small $\lvert \Rv \rvert$ are
\begin{align}
G^{xx}(1,0) &= -\rho_{x}^{2} \\
G^{xx}(0,1) &= \alpha^{2}\rho_{y}^{2} \\
G^{xx}(1,1) &= \rho_{y}\left(\rho_{x} - \frac{\alpha}{\pi}\right) \\
G^{xx}(2,1) &= \left(\frac{\rho_{x}}{\alpha} - \frac{1}{\pi}\right)^{2} \\
G^{xx}(0,2) &= -\left(\frac{\alpha}{\pi}-\alpha^{2}\rho_{y}\right)^{2} \\
G^{xx}(0,3) &= \alpha^{2}\left[\rho_{y}(1+2\alpha ^{2})-\frac{2\alpha}{\pi}\right]^{2} \punc ,
\end{align}
while the asymptotic behavior for $\lvert \Rv \rvert \gg 1$ is algebraic, rather than exponential:
\begin{multline}
\frac{G^{xx}(\Rv)}{\alpha^{2}} \approx (-1)^{X}\frac{1}{\pi^{2}[X^2 + (\alpha Y)^{2}]^{2}} \times {} \\ 
\begin{cases}
X^{2} & \text{for \(X\) odd, $Y$ even} \\
(\alpha Y)^{2} & \text{\phantom{for} \(X\) even, $Y$ odd} \\
(\alpha Y)^{2}& \text{\phantom{for} \(X\) odd, $Y$ odd} \\
X^{2}-1 & \text{\phantom{for} \(X\) even, $Y$ even.}
\end{cases}
\end{multline}

Similarly, the connected correlation function of two vertical dimers with separation $\Rv$, illustrated in \reffig{fig:ddcf} (bottom left), is
\begin{equation}
\label{eq:gyy}
G^{yy}(\Rv) = \langle d_{j+X,y}(l+Y)d_{j,y}(l)\rangle - \langle d_{j+X,y}(l+Y)\rangle\langle d_{j,y}(l)\rangle \punc.
\end{equation}
Following the same procedure as for $G^{xx}(\bm{R})$, but now using Eqs.~(\ref{eq:dyc}) and (\ref{eq:cc})--(\ref{eq:ccd}), yields
\begin{equation}
G^{yy}(\Rv) =
\begin{cases}
\Gamma(\Rv)^{2} & \text{for \(X+Y\) odd} \\
\Delta(\Rv)^{2} - \Gamma(\Rv)^{2} & \text{\phantom{for} \(X+Y\) even.}
\end{cases}
\end{equation}
Note that the correlators \(G^{xx}\) and \(G^{yy}\) should be related by \(90^\circ\) rotations, in a similar way to that shown for \(Z(\bm{t})\) in \refsec{partitionfunction}.

The third possibility is the connected correlation function of a horizontal and vertical dimer with separation $\Rv$, illustrated in \reffig{fig:ddcf} (bottom right), which is
\begin{equation}
\label{eq:gxy}
G^{xy}(\Rv) = \langle d_{j+X,y}(l+Y)d_{j,x}(l)\rangle - \langle d_{j+X,y}(l+Y)\rangle\langle d_{j,x}(l)\rangle \punc .
\end{equation}
The result is
\begin{multline}
\frac{G^{xy}(\Rv)}{\alpha} = {} \\
\begin{cases}
\Gamma(\Rv)[\Delta(X-1,Y) - \Gamma(X-1,Y)] & \text{for \(X+Y\) odd} \\
\Gamma(X-1,Y)[\Gamma(\Rv)-\Delta(\Rv) ] & \text{\phantom{for} \(X+Y\) even,}
\end{cases}
\end{multline}
with asymptotic behavior
\begin{multline}
\frac{G^{xy}(\Rv)}{\alpha} \approx (-1)^{X+Y}\frac{1}{\pi^{2}[X^2 + (\alpha Y)^{2}]^{2}} \times {} \\
\begin{cases}
X\alpha (Y+1) & \text{for \(X\) odd, $Y$ even} \\
(X-1)\alpha Y & \text{\phantom{for} \(X\) even, $Y$ odd} \\
X\alpha Y & \text{\phantom{for} \(X\) odd, $Y$ odd} \\
(X-1)\alpha (Y+1) & \text{\phantom{for} \(X\) even, $Y$ even.}
\end{cases}
\end{multline}
The results in this section are in agreement with Sec.~7 of \refcite{Fisher1963}.

\subsection{Monomer distribution function}
\label{monomerdistributionfunction}

Finally, we characterize the (entropic) interaction between a pair of inserted test monomers by calculating the monomer distribution function
\beq[EqGm]
G\sub{m}(\Rv) = \frac{1}{Z(\bm{t})}\sum_{c\in\goC(\rv_+,\rv_-)} \alpha^{N_{x}}\punc ,
\eeq
where the set \(\goC(\rv_+,\rv_-)\) contains all configurations with monomers at sites \(\rv_\pm\). For simplicity, we consider the case of two monomers on the same row, though the formalism can be extended to the general case.

Because $\sigma_{j}^{-}$ inserts a monomer on site $j$, in the transfer-matrix formalism one has
\begin{equation}
\label{eq:gmspin}
G\sub{m}(X,0) = \langle \sigma_{j}^{-}(l)\sigma_{j+X}^{-}(l) \rangle \punc, 
\end{equation}
which becomes
\begin{equation}
\label{eq:gm2}
G\sub{m}(X,0) = -\left\langle C_{j} \left[\prod_{i = j + 1}^{j + X - 1}(1 - 2C_{i}^{\dagger}C_{i})\right]C_{j+X} \right\rangle
\end{equation}
after performing the Jordan--Wigner transformation, \refeqqs{eq:JW1}{eq:JW3} (from here on we do not explicitly show dependence on the row $l$).\footnote{In the case of two monomers on different rows, the operator on each row has an odd number of $C_{j}$ operators and so does not commute with \((-1)^{\Phi_y}\). To treat this case, we would not be able to use \refeqq{eq:Qexpval} and would instead require the analogous expression for \(O\) anticommuting with \((-1)^{\Phi_y}\).}

Following \refcites{Sachdev2011, Lieb1961, Schultz1964}, we now define operators
\begin{align}
A_{j} &= C_{j}^{\dagger} + C_{j} \\
B_{j} &= C_{j}^{\dagger} - C_{j}
\end{align}
(note that $1 - 2C_{j}^{\dagger}C_{j} = A_{j}B_{j}$), which, by \refeqqs{eq:cc}{eq:ccd}, satisfy
\begin{align}
\label{eq:aa}
\langle A_{j}A_{j+X} \rangle &= \delta_{X,0} \\
\label{eq:bb}
\langle B_{j}B_{j+X} \rangle &= -\delta_{X,0} \\
\label{eq:ba}
\langle B_{j}A_{j+X} \rangle &= -\langle A_{j+X}B_{j} \rangle = -2\Gamma(X,0) \punc .
\end{align}
In terms of these, \refeqq{eq:gm2} is a sum of four $2X$-point correlators, each of which can be expressed as a sum of products of two-point correlators through Wick's theorem \cite{Abrikosov2012}. Then, by \refeqqand{eq:aa}{eq:bb}, the two correlators containing an unequal number of $A_{j}$ and $B_{j}$ vanish, while the remaining two are
\begin{align}
\mathcal{W}(B,A) &= \frac{1}{4}\left\langle \prod_{i=j}^{j+X-1}B_{i}A_{i+1} \right\rangle \\ &= \frac{1}{4}\sum_{\sigma\in S_{X}}\sgn(\sigma)\prod_{i=1}^{X}\langle B_{j+i-1}A_{j+\sigma_{i}} \rangle \punc ,
\end{align}
where $S_{X}$ denotes the symmetric group of order $X$, and $(-1)^{X-1}\mathcal{W}(A,B)$. Inserting \refeqq{eq:ba} and using the relation $\Gamma(-X,0) = (-1)^{X}\Gamma(X,0)$ with $\prod_{i=1}^{X}(-1)^{i - \sigma_{i}} = 1$, it follows that $\mathcal{W}(A,B)=\mathcal{W}(B,A)$, and hence
\begin{multline}
G\sub{m}(X,0) = {} \\
\begin{cases}
\displaystyle \frac{1}{2}\sum_{\sigma\in S_{X}}\sgn(\sigma)\prod_{j=1}^{X} -2\Gamma(1-(j-\sigma_{j}),0) & \text{for \(X\) odd}\\
0 & \text{\phantom{for} \(X\) even,}
\end{cases}
\end{multline}
which can be expressed as a Toeplitz determinant
\beq
G\sub{m}(X,0) = \frac{1}{2}\det T_X \qquad \text{for $X$ odd,}
\eeq
where \(T_X\) is an \(X\times X\) matrix with elements \((T_X)_{j,j'} = -2\Gamma(1-(j-j'),0)\).

From \reftab{tab:gvalues}, the first two non-zero values are
\begin{align}
G\sub{m}(1,0) &= \frac{\rho_{x}}{\alpha} \\
G\sub{m}(3,0) &= \frac{4\rho_{x}}{\alpha^{5}} \left[(1 + \alpha^{2})^{2} \rho_{x}^{2} - \frac{\alpha^{2}}{\pi^{2}}\right]
\end{align}
[cf. Eqs.~(11.1) and (11.3) of \refcite{Fisher1963}], where, up to a factor of $\alpha$, the former is equivalent to the occupation probability of a horizontal bond as calculated in \refsec{dimeroccupationnumber}. 

To calculate the asymptotic behavior for large \(X\), we define \(\varphi(k) = -2\sum_{j = -\infty}^\infty \ee^{\ii k j}\Gamma(1-j,0) = -\ee^{\ii k}\ee^{2\ii \theta_k}\sgn(k)\) for \(-\pi \le k < \pi\). Unlike on the triangular lattice \cite{Fendley2002,Basor2006}, Szeg\H{o}'s limit theorems do not apply, since \(\varphi\) is not a continuous function, and instead we apply the Fisher--Hartwig conjecture \cite{Fisher1969}.

The discontinuities at \(k = 0\) and \(k = \pm \pi\) can be expressed by defining \(t_{\beta}(k) = \ee^{-\ii \beta (\pi - k)}\) for \(0 < k < 2\pi\) \cite{Basor1991}, in terms of which \(\varphi(k) = b(k)t_{1/2}(k)t_{1/2}(k-\pi)\). Here, \(b(k) = -\ii \ee^{2\ii \theta_k}\) is continuous and has zero winding number when viewed as a map from \(\ee^{\ii k}\) to the unit circle. Its Wiener--Hopf factorization, \(b(k) = b_+(\ee^{\ii k})b_-(\ee^{\ii k})\), with \(b_+\) (\(b_-\)) analytic and nonzero everywhere inside (outside) the unit circle \cite{Bottcher2006}, is
\beq
b_{\pm}(z) = \sqrt{\pm\frac{c_\pm - z}{c_\pm + z}}\punc,
\eeq
where \(c_{\pm} = \alpha^{-1} \pm \sqrt{1 + \alpha^{-2}}\).

According to the Fisher--Hartwig conjecture \cite{Basor1991}, we then have
\beq
\det T_X \approx G[b]^{X} X^\Omega E\punc,
\eeq
for large \(X\), with \(G[b]=1\), \(\Omega = -\frac{1}{2}\) and
\beq
E = \frac{2^{2/3}\ee^{6\zeta'(-1)}}{(1+\alpha^2)^{1/4}} \simeq \left(\frac{1+\alpha^2}{2}\right)^{-1/4}\times 0.494744 \punc,
\eeq
where \(\zeta'\) is the derivative of the Riemann zeta function.

The monomer distribution function therefore obeys
\beq
G\sub{m}(X,0) \approx \frac{E}{2\sqrt{X}} \qquad \text{for $X\gg 1$, odd.}
\eeq
A consistent result was found by Hartwig \cite{Hartwig1966} for the case of monomers separated along a diagonal (i.e., \(X=Y\)) using the Pfaffian method.

Note that the algebraic dependence on \(X\), stemming mathematically from the discontinuity in \(\varphi\), contrasts with the exponential behavior on the triangular lattice \cite{Fendley2002,Basor2006}. As noted by Au-Yang and Perk \cite{AuYang1984}, the decrease with \(X^{-1/2}\) can be understood by relating the dimer model to two uncoupled Ising models at the critical point.

\section{Conclusions}
\label{conclusions}

We have expressed Lieb's transfer matrix for the classical square-lattice dimer model in terms of a free-fermion Hamiltonian, and used its spectrum to rederive some useful results. Although these can equally be derived using Pfaffian techniques, the second quantized approach presented in this paper is perhaps more elegant.

Specifically, our results include the torus partition function which, by including a field $\bm{t}$, can be interpreted as a moment-generating function of the flux. We have also shown how expectation values can be expressed in terms of the fermionic operators, and evaluated dimer occupation numbers, dimer--dimer correlation functions and the monomer distribution function in the thermodynamic limit, all of which are independent of flux sector for not-too-large flux. Finally, we have derived a new result, namely the asymptotic behavior of the monomer distribution function for large monomer separation along the same row.

The results in this paper are also relevant to the corresponding quantum dimer model at its Rokhsar--Kivelson point \cite{Rokhsar1988}, while the transfer-matrix method can be extended to other two-dimensional lattices. Indeed, the straightforward generalization of Lieb's transfer matrix to the (bipartite) honeycomb and square-octagon lattices, which can both be viewed as a square lattice with certain horizontal bonds removed [i.e., certain terms omitted from the sum in $V_{3}$; see \refeqq{eq:V3}], has already been demonstrated in \refcite{Grande2011}.

One advantage of the transfer-matrix method is that dimer--dimer interactions can be easily included in the operator formalism, in terms of products of the dimer occupation numbers $d_{j,x}$ and $d_{j,y}$. For example, on a row of vertical bonds, the operator $\sum_{j}d_{j,y}d_{j+1,y}$ describes interactions between parallel pairs of nearest-neighbor dimers, as studied in \refcites{Alet2005,Alet2006b}. This is a four-fermion interaction, which is non-integrable \cite{Alet2006b} but could be included perturbatively using standard diagrammatic perturbation theory.

Furthermore, as we will show in a forthcoming publication \cite{BosonizationPaper}, the well-known height field theory \cite{Blote1982,Zeng1997} of the two-dimensional classical dimer model can be rigorously derived from the fermionic Hamiltonian of \refeqq{eq:ham2}. This can be achieved by taking a long-wavelength limit and using the technique of bosonization \cite{vonDelft1998} to express the theory in terms of a single free bosonic field. Interaction operators included perturbatively in this context manifest themselves through renormalization of the `stiffness' as well as the introduction of (cosine) potential terms consistent with symmetry requirements \cite{BosonizationPaper}.


\appendix

\section{Jacobi theta functions}
\label{jacobithetafunctions}

We define the Dedekind eta function
\begin{equation}
\label{eq:dedekind}
\eta (q) = q^{1/24}\prod_{n=1}^{\infty}(1 - q^{n}) \punc ,
\end{equation}
for nome $q$ such that $\lvert q \rvert < 1$. For a complex number $y$, the Jacobi theta functions are
\begin{align}
\begin{split}
\label{eq:t1}
\theta_{1}(y\vert q) &= -\ii\sqrt{y} q^{1/12}\eta(q)\prod_{n=1}^{\infty}(1-yq^n)(1-y^{-1}q^{n-1}) \\ &= -\ii\sum_{r\in \mathbb{Z} + 1/2} (-1)^{r-1/2}y^rq^{r^2/2}
\end{split} \\
\begin{split}
\label{eq:t2}
\theta_{2}(y\vert q) &= \sqrt{y} q^{1/12}\eta(q)\prod_{n=1}^{\infty}(1+yq^n)(1+y^{-1}q^{n-1}) \\ &= \sum_{r\in \mathbb{Z} + 1/2} y^rq^{r^2/2}
\end{split} \\
\begin{split}
\label{eq:t3}
\theta_{3}(y\vert q) &= q^{-1/24}\eta(q)\prod_{n=1}^{\infty}(1+yq^{n-1/2})(1+y^{-1}q^{n-1/2}) \\ &= \sum_{n\in \mathbb{Z}} y^nq^{n^2/2}
\end{split} \\
\begin{split}
\label{eq:t4}
\theta_{4}(y\vert q) &= q^{-1/24}\eta(q)\prod_{n=1}^{\infty}(1-yq^{n-1/2})(1-y^{-1}q^{n-1/2}) \\ &= \sum_{n\in \mathbb{Z}} (-1)^{n}y^nq^{n^2/2} \punc .
\end{split}
\end{align}
In terms of these definitions, which follow \refcite{Rasmussen2012}, the functions defined in Section 20 of \refcite{NIST:DLMF} are \(\theta^{\text{NIST}}_i(z,q) = \theta_i(\ee^{2\ii z}|q^2)\).

These functions obey the modular identities \cite[Sec.~20.7]{NIST:DLMF}
\beq[eq:modularidentity]
\thetav(\ee^s|\ee^{-2\pi\rho}) = \frac{1}{\sqrt{\rho}} \ee^{\frac{s^2}{4\pi \rho}}\begin{pmatrix}
-\ii&0&0&0\\
0&0&0&1\\
0&0&1&0\\
0&1&0&0
\end{pmatrix}\thetav(\ee^{\ii s/\rho}|\ee^{-2\pi/\rho}) \punc ,
\eeq
where \(\thetav(y|q) =  \big(\theta_1(y|q),\dotsc, \theta_4(y|q)\big)^T\), and \cite[Sec.~23.18]{NIST:DLMF}
\beq[eq:modularidentityrho]
\eta(\ee^{-2\pi \rho}) = \frac{1}{\sqrt{\rho}}\eta(\ee^{-2\pi/\rho})\punc.
\eeq

\bibliography{dimersbibliography}

\end{document}